\documentclass[a4paper,11pt]{article}
\pdfoutput=1

\usepackage{jheppub} 
                     
\usepackage{amsmath}
\usepackage[T1]{fontenc} 
\usepackage{mathtools}
\usepackage{xcolor}
\newcommand{\dd}{\mathrm{d}}

\usepackage[normalem]{ulem}

\title{How to find regular black-hole islands in the swamp}

\author[a]{Astrid Eichhorn,} 
\author[b,a]{Pedro G.~S.~Fernandes,}
\author[a]{Verena Koeder}

\affiliation[a]{Institut f\"ur Theoretische Physik, Universit\"at Heidelberg, Philosophenweg 12 \& 16, 69120 Heidelberg, Germany}
\affiliation[b]{CENTRA, Departamento de F\'{\i}sica, Instituto Superior T\'ecnico -- IST, Universidade de Lisboa -- UL,
Avenida Rovisco Pais 1, 1049-001 Lisboa, Portugal}

\emailAdd{eichhorn@thphys.uni-heidelberg.de}
\emailAdd{pedrogsfernandes@tecnico.ulisboa.pt}
\emailAdd{koeder@thphys.uni-heidelberg.de}

\abstract{An infinite tower of higher-curvature corrections is expected to lead to regular black-hole solutions. We assume that this regular behavior does not abruptly arise, but gradually builds up with each successive order of the effective field theory of gravity. In addition, we assume that the effective field theory comes with a single scale of singularity-resolving new physics. We translate this assumption into a monotonicity requirement for deviations of the metric functions from a Schwarzschild black hole. Under these assumptions, conditions can be formulated which trace the increase of ``less singular'' black-hole behavior on and outside of the black-hole horizon within the regime of validity of the effective field theory. These conditions constrain the effective field theory coefficients. To exemplify this construction, we motivate five specific conditions and find a corresponding island of black-hole regularity conditions at order eight in the effective field theory. We also explore whether string theory as well as asymptotically safe gravity lie on this island.}

\begin{document}
\maketitle
\newpage

\section{Introduction}
We have ample evidence for the existence of black holes~\cite{LIGOScientific:2016aoc,LIGOScientific:2016sjg,LIGOScientific:2017bnn,LIGOScientific:2017vox,LIGOScientific:2017ycc,LIGOScientific:2017vwq,EventHorizonTelescope:2019dse,EventHorizonTelescope:2022wkp,EventHorizonTelescope:2022xqj}. Yet, we currently lack a proper understanding of their true nature. 
In General Relativity (GR), individual black holes in vacuum are described by the Kerr solution, with small modifications due to surrounding matter, e.g., accretion disks. After a binary merger of black holes  or neutron stars, or the gravitational collapse of a compact star, GR predicts that the resulting spacetime geometry settles down to a Kerr geometry.
However, we know that the Kerr geometry does not fully describe astrophysical black holes, because it breaks down not just at curvature singularities, but at non-zero spin already at the Cauchy horizon.   
If the observed astrophysical black holes can be described by a spacetime metric, they should not contain a curvature singularity anywhere, should be geodesically complete, and should not contain any Cauchy horizons or other hindrances to a deterministic physical picture. We call such objects \emph{regular black holes}, first proposed in \cite{Bardeen1968}, see \cite{Carballo-Rubio:2025fnc} for a recent review.

Recent progress towards a regular-black-hole description of these objects has been achieved along multiple directions: for instance, in \cite{Eichhorn:2025pgy}, a modified-gravity theory has been found in which black-hole solutions are regular for any value of the mass.
A symmetry-reduced setting, in which dynamical questions can be studied has been developed in \cite{Carballo-Rubio:2025ntd} and applied, e.g., in \cite{Boyanov:2025pes, Borissova:2026wmn,Carballo-Rubio:2026mvj, Borissova:2026prd,Carballo-Rubio:2026mns}.
Further, spherically symmetric solutions of infinite-order curvature theories without curvature singularities were found in \cite{Bueno:2025zaj}, see also \cite{Bueno:2024dgm,Bueno:2024zsx,Bueno:2024eig,Bueno:2025gjg,Bueno:2026oyg,Bueno:2026wjt,Fernandes:2025fnz,Fernandes:2025eoc,Fernandes:2025mic} for other settings. The intuition that some form of non-locality is required to achieve regularity is also supported by \cite{Borissova:2025nvj}.
Finally, in \cite{DiFilippo:2025kzh,Gralla:2025gzl} it was argued that time-dependence introduced by Hawking radiation may suffice to avoid the formation of a singularity.

Nevertheless, a fully satisfactory description of a regular black-hole spacetime has not been achieved yet, see, e.g., \cite{Carballo-Rubio:2025fnc} for a discussion of outstanding challenges. The above examples typically fall short on two counts: First, various instabilities suggest that understanding the time-dependence of these geometries may be crucial \cite{Carballo-Rubio:2025fnc}. 
Second, these developments are typically not well grounded in fundamental theory. Rather than starting from a candidate quantum gravity theory, the above developments are limited to either semi-classical effects, a truncation of the full effective action from quantum gravity or an ad-hoc ``guess'' of an effective action, or even just limited to the kinematical level.

In the present work, we aim at addressing the second shortcoming. Rather than starting from a particular framework for quantum gravity, we start from a point of common ground between all quantum-gravity frameworks. It is generally agreed upon that a viable theory of quantum gravity must reduce, at low curvatures, to GR corrected by higher-order curvature terms.\footnote{The full effective action arising from integrating out massless graviton fluctuations is also expected to contain non-localities which are more challenging to treat, e.g., logarithmic terms. We assume that for black-hole spacetimes, such terms are effectively negligible and only become dynamically important at much larger distance scales, e.g., in the deep infrared regime of cosmology.} 
Differences between quantum-gravity frameworks, if they exist at this level, manifest in different values of the effective-field-theory (EFT) coefficients. However, it is in fact not even clear to what extent physical differences exist, or whether distinct approaches share universal physical features \cite{deBoer:2022zka}. Against this background, it is a highly relevant question to understand how black-hole regularity may manifest itself in the effective field theory\footnote{See, e.g., Refs.~\cite{deRham:2020ejn,Endlich:2017tqa,Melville:2024zjq,Horowitz:2023xyl,Horowitz:2024dch,deRham:2021bll,deRham:2022gfe,Bern:2021ppb,Reall:2019sah,Bernard:2025dyh,Cardoso:2018ptl,deRham:2019ctd,Figueras:2024bba,Cano:2020cao,Cano:2021myl,Cano:2022wwo,Cano:2023tmv,Cano:2023jbk,Cano:2024wzo,Cano:2024bhh,Cano:2024jkd,Cano:2024ezp,Maenaut:2024oci,Cano:2025mht,Cano:2025ejw,vanderSteen:2026qva,Fernandes:2025vxg,Hu:2026ute,Boyce:2026rnn} for works that explore black holes in the context of the EFT of gravity.}.

We assume that regularity of a black-hole solution does not abruptly arise, once an infinite tower of higher-curvature corrections has been accounted for. Rather, we expect that the black-hole singularity, as well as further problematic features of black-hole spacetimes, are gradually weakened as successive orders in the EFT are accounted for.
Thus, we formulate a set of conditions that largely relate to properties of the black-hole horizon and that we consider \emph{tracers} of an approach to regularity.

We explore the resulting constraints on the effective action from demanding that all tracers show the onset of regular behavior. This may be interpreted as delineating a \emph{landscape} of quantum gravity which contains regular black holes with particular properties, from a \emph{swampland} which does not. We do, however, not claim that theories in this \emph{swampland} are necessarily excluded. Until we have a working and observationally well-supported quantum theory of gravity, we should not exclude the possibility that conditions that we consider as perfectly reasonable are not realized in Nature. Nevertheless, we consider the resulting \emph{island} of effective field theories preferred, because we expect that it may realize the most straightforward approach towards regularity of black-hole solutions.\newline

This paper is structured as follows: In Sec.~\ref{sec:conjectures}, we motivate tracers of black-hole regularity that are accessible to the EFT. In Sec.~\ref{sec:field_redefinitions}, we caution against the use of field redefinitions in EFT analysis, because field redefinitions may result in non-minimal matter couplings, altering properties of black holes that pertain to propagating matter (e.g., the causal structure that matter ``sees'').
We introduce the EFT up to order eight in derivatives in Sec.~\ref{sec: EFT} and then constrain the EFT coefficients using the tracers of black-hole regularity in Sec.~\ref{sec: EFT results}. In Sec.~\ref{sec: EFT with scalar field}, we repeat this analysis in an EFT containing a light scalar field in addition to gravity. 
We then confront the EFT constraints with results for the EFT coefficients from string theory and asymptotically safe gravity in Sec.~\ref{sec: comparison to ST and ASQG}. We draw conclusions and provide an outlook in Sec.~\ref{sec: conclusion}.
Further technical details are provided in the appendix.

\section{Conjectures on the properties of regular black holes in quantum gravity}\label{sec:conjectures}
We expect that infinite towers of corrections to the Einstein-Hilbert action cure the singularity problem in black-hole spacetimes. The onset of regularity is very likely not abrupt, but appropriate tracers of regularity should gradually increase order-by-order in the curvature expansion. 
To formulate appropriate tracers, we note that the EFT expansion breaks down at $r=0$, because all nonzero higher-order curvature terms diverge at this point for a static and spherically-symmetric vacuum solution of GR. Thus, the appropriate tracers of regularity that we will use pertain to features of the spacetime \emph{away} from the singularity. In particular, we expect that specific properties of the horizon and other distinct surfaces in black-hole spacetimes are suitable tracers.

This expectation critically hinges on a monotonicity assumption: we assume that an effective \emph{weakening} of gravity, as is needed to resolve singularities and results in deviations from the Kerr spacetime, increases monotonically throughout the spacetime from asymptotic infinity all the way to the core. This can be understood as the assumption that 
new physics that weakens gravity is parameterized by a single scale~\cite{Eichhorn:2022oma}.
Then, a (quantitatively small) imprint of this effect can be found, e.g., at the horizon.
It is not logically excluded that this monotonicity assumption does not hold and a regular black-hole solution in quantum gravity features, e.g., deviations from the Kerr spacetime with different signs at different locations throughout the spacetime. We consider this unlikely, as it makes the structure of deviations from the Kerr spacetime not simple in the sense that their physical effect (weakening vs. strengthening of gravity) is different at different locations in the spacetime (corresponding to different curvature scales). We know of no convincing argument that would support such a behavior.
We spell out the resulting simplicity assumption in detail below.

To formulate our tracer conditions we focus on static and spherically symmetric black holes. Without loss of generality, their metrics can be written as
\begin{equation}
\begin{aligned}
    \dd s^2 &=- h(r) \dd t^2 + \frac{1}{f(r)}\, \dd r^2 + r^2 \mathrm{d} \Omega^2\\
    &=- \left( 1-\frac{r_S}{r}f_1(r)\right)\, \dd t^2 + \left( 1-\frac{r_S}{r}f_2(r)\right)^{-1}\, \dd r^2 + r^2 \dd \Omega_2
    \label{eq:metric}
    \end{aligned}
\end{equation}
where $f_{1,2}$ are two $r$-dependent functions, and $r_S=2M$ is the Schwarzschild radius of the black hole. The actual location of the event horizon in this spacetime is shifted away from $r=r_S$, if $f_{1,2}(r)\neq 1$. Thus, $(f_{1,2}(r)-1)$ encodes the deviation from Schwarzschild.

Below, we derive constraints on the EFT coefficients by requiring that the metric Eq.~\eqref{eq:metric} satisfies particular conditions, that link regularity to properties of the black-hole spacetime away from $r=0$. These conditions are distinct ways of quantifying the expectation that a less singular black hole spacetime exhibits ``weaker gravity''.

In our analysis below we assume that minimally coupled matter follows the geodesics of this metric, or, put differently, the matter Lagrangian is constructed with this metric. This is the natural assumption to make in our setting, where we consider the low-energy effective field theory (EFT) that arises after high-energy degrees of freedom have been integrated out. This implies that in the EFT, we must not perform any field redefinition for the metric, without accounting for the resulting change in the coupling to matter. 
\subsection{Simplicity and regularity increase the compactness of the horizon and photon sphere}
Regularization of curvature singularities is a priori independent of the properties of the spacetime at larger distances from the black-hole core. However, regularity has immediate implications for all locations in a black-hole spacetime, if it is combined with a simplicity assumption. From an EFT perspective, this simplicity assumption can be motivated from a naturalness argument which says that all new-physics effects are suppressed by a single new-physics scale. Thus, the following three assumptions have been put forward in \cite{Eichhorn:2022oma} as the most natural expectation for physically viable, regular black holes:
\begin{enumerate}
    \item The UV completion of gravity, whatever it is, is responsible for curing the black-hole singularity, such that no curvature scalars diverge anywhere (\textit{regularity}).
    \item The regular black-hole metric realized by the UV completion exhibits the standard Newtonian limit at large distances (\textit{asymptotic flatness}).
    \item Deviations from the Schwarzschild spacetime are monotonic functions of the curvature scale and characterized by a single new length scale associated with the UV completion (\textit{simplicity}) \cite{Eichhorn:2022oma}.
\end{enumerate}
Let us apply these conditions to Eq.~\eqref{eq:metric} one after another. Regularity entails 
\begin{equation}
f_{1,2}\rightarrow \mathcal{O}(r^3)\quad \mbox{as } r \rightarrow 0. 
\end{equation}
 
The correct Newtonian limit
requires 
\begin{equation}
f_{1,2}(r) \rightarrow 1\quad \mbox{for } r \rightarrow \infty, 
\end{equation}
while simplicity results in the monotonicity requirement
\begin{equation}
f'_{1,2}(r)>0,\quad \forall r>0. 
\end{equation}
Putting all conditions together, we find \cite{Eichhorn:2022oma}
\begin{equation}
    0< f_{1,2}(r)< 1,\qquad f'_{1,2}(r)>0,
\end{equation}
for all values $0<r<\infty$.

Within a regular spacetime characterized by these assumptions, we can make statements about two physical, special hypersurfaces of the spacetime, namely the horizon and the photon sphere of a black hole. 
The event horizon is located at $r=r_h$ such that $g^{rr}(r_h)=0$, leading to
\begin{equation}
    r_h = r_Sf_2(r_h),
\end{equation}
and since $0<f_2(r_h)<1$, we find
\begin{equation}
    r_h<r_S.
    \label{eq:rh_cond}
\end{equation}
This means that, just by simplicity, the horizon of a regular black-hole is more compact than that of its singular counterpart.

In fact, the same result applies to the photon sphere, the innermost stable circular orbit, or any other special surface in the spacetime. We focus on the photon sphere here, due to its relevance for black-hole images.
Its radial location $r_{\gamma}$ is determined by 
\begin{equation}
h(r_{\gamma}) - \frac{r_{\gamma}}{2} h'(r_{\gamma})=0, \label{eq:condition_photonsphere}
\end{equation}
which results in
\begin{equation}
    r_\gamma=\frac{3 M f_1\left(r_\gamma\right)}{1+M f_1^{\prime}\left(r_\gamma\right)}<\frac{3 M}{1+M f_1^{\prime}\left(r_\gamma\right)}<3 M\,.
\end{equation}
Just like for the case of the horizon, the photon sphere of a regular black hole is more compact than that of a Schwarzschild black hole.
Physically, this can be understood as follows: In order for singularity resolution to occur, gravity needs to weaken at the core of a black hole. Requiring that the corresponding deviation from GR does not introduce more than one scale, there must be a weakening of gravity at all distances (that itself becomes a progressively smaller effect as the distance to the black-hole core grows). Distinct surfaces within a black-hole spacetime are all linked to gravity becoming \emph{strong} in particular ways. For instance, at the photon sphere, gravity causes sufficiently strong lensing for photon orbits to close. At the horizon, gravity is strong enough to render the expansion of outgoing null geodesics zero, rather than positive. Weakening gravity everywhere throughout the spacetime therefore implies that all these distinct surfaces have to become more compact, as the strength of gravity required for them to form is only reached closer to the center of the black hole.

\subsection{Reduced spaghettification}
Our next conjecture related to the manifestation of gravity becoming weaker is that of decreased tidal forces at the event horizon. This conjecture hinges on the fact that diverging curvature is related to diverging tidal forces. We assume that the tidal forces at the horizon are reduced.  

The geodesic deviation equation describes how the space-like separation $\eta$ of two geodesics evolves in proper time
\begin{equation}
\frac{D^2 \eta^{\hat{\alpha}}}{D \tau^2}=K^{\hat{\alpha}}{ }_{\hat{\beta}} \eta^{\hat{\beta}}\,,
\end{equation}
where, following \cite{Lima:2020wcb, Crispino:2016pnv}, the tidal tensor is calculated using the tetrad frame of an observer following a radial geodesic
\begin{equation}
K^{\hat{\alpha}}{ }_{\hat{\beta}}=R^\eta{ }_{\nu \gamma \delta} e^{\hat{\alpha}}{ }_\eta e^\nu{ }_{\hat{0}} e^\gamma{ }_{\hat{0}} e^\delta{ }_{\hat{\beta}}\,,
\end{equation}
where the hatted indices belong to the tetrad basis and the non-hatted ones to the coordinate basis.
For a general spherically symmetric and static spacetime as in Eq.~\eqref{eq:metric}, the tetrad of a radially infalling observer is
\begin{equation}
\begin{aligned}
e_{\hat{0}}^\mu & =\left(\frac{E}{h(r)},-\sqrt{f(r)\left(\frac{E^2}{h(r)}-1 \right)}, 0,0\right), \\
e_{\hat{1}}^\mu & =\left(-\frac{\sqrt{E^2-h(r)}}{h(r)}, E \sqrt{\frac{f(r)}{h(r)}}, 0,0\right), \\
e_{\hat{2}}^\mu & =\left(0,0, r^{-1}, 0\right), \\
e_{\hat{3}}^\mu & =\left(0,0,0,(r \sin \theta)^{-1}\right) \,,
\end{aligned}
\end{equation}
where $E$ is the energy of the observer. 
The non-vanishing components of the tidal tensor in this tetrad basis read
\begin{equation}
\begin{aligned}
& K_{\hat{r} \hat{r}}=\frac{f(r) h'(r)^2-h(r) \left(f'(r) h'(r)+2 f(r) h''(r)\right)}{4 h(r)^2}\,, \\
& K_{\hat{\theta} \hat{\theta}}=\frac{h(r) \left(E^2-h(r)\right) f'(r)-E^2 f(r) h'(r)}{2 r h(r)^2}, \\
& K_{\hat{\phi} \hat{\phi}}=\frac{h(r) \left(E^2-h(r)\right) f'(r)-E^2 f(r) h'(r)}{2 r h(r)^2}\,.
\end{aligned}
\end{equation}
The radial component encodes how the observer is either ``spaghettified'' or compressed radially as they move towards the black hole.
This component is also the only one that does not depend on the energy of the  
observer, enabling us to make a statement that only depends on the spacetime geometry, but not on the specific properties of the observer.

\subsection{Proper time of infalling observers}
Physically, diverging tidal forces are just one part of the problem of black holes in GR. The other consists in geodesic incompleteness, i.e., the finite amount of proper time it takes for an infalling observer to reach $r=0$, which is strictly speaking not a point within the spacetime.
Therefore, we conjecture that, at finite order in the EFT expansion, the amount of proper time that it takes an infalling observer to reach a given location in the spacetime should be extended. Additional tracers of regularity are linked to other types of motion, but we focus on radial geodesic motion here.

For the general spherically symmetric metric in Eq.~\eqref{eq:metric}, a time-like geodesic of a massive particle with energy $E$ and angular momentum $L$ satisfies
\begin{equation}
    \mathrm{d}\tau=\pm \frac{\mathrm{d}r}{\sqrt{-f(r)\left(1+\frac{L^2}{r^2}-\frac{E^2}{h(r)} \right)}}\,,
    \label{eq: proper time int}
\end{equation}
where $E^2=h(r)+\frac{h(r)}{f(r)}\Dot{r}^2$ is constant, and $\dot r \equiv \dd r/\dd \tau$. For an observer initially at rest at $r=R$, we obtain $E^2=h(R)$. We constrain the EFT solely from radial motion, which corresponds to $L=0$.

The previous discussion of the simplicity property has shown that both the photon sphere and the horizon are more compact in regular black holes when compared to their  counterparts in a singular black-hole spacetime. Analyzing the proper time it takes for a radially falling observer to travel between photon sphere and horizon is useful in order to understand how these surfaces become more compact \textit{in relation to one another}. Even for a time-like observer, the photon-sphere radius has a significant meaning, as it influences the trajectories of light signals the observer might send. We therefore consider the proper time perceived by a time-like observer initially  hovering at the photon sphere and then proceeding to a free-fall towards the black-hole horizon. \\
For a general spherically symmetric static metric that includes modifications to the Schwarz- schild metric proportional to a small parameter $\varepsilon$, the corresponding proper time relation is 
\begin{equation}
    \Delta \tau_{r_\gamma \rightarrow\,r_h}=\int_{r_h(\varepsilon)}^{r_\gamma(\varepsilon)}\frac{\mathrm{d}r}{\sqrt{-f(r,\,\varepsilon)\left(1-\frac{h(r_\gamma(\varepsilon),\,\varepsilon))}{h(r,\,\varepsilon)} \right)}}\,.
\end{equation}
Series expansion under the integral requires a linear coordinate rescaling
\begin{equation}
    r(x)= r_h(\varepsilon)+x ( r_\gamma(\varepsilon)-r_h(\varepsilon))\,,
\end{equation}
with which the integral boundaries become independent of $\varepsilon$
\begin{equation}
    \Delta \tau_{r_\gamma \rightarrow\,r_h}=\int_{0}^{1}\mathrm{d}x\frac{\left(r_\gamma(\varepsilon)-r_h(\varepsilon) \right)}{\sqrt{-f(x,\,\varepsilon)\left(1-\frac{h(1,\,\varepsilon)}{h(x,\,\varepsilon)} \right)}}\,.
\end{equation}
The expression under the square root has a root at $x=1$ and as its derivative is non-zero at $x=1$, this is a simple root. We can now isolate it from the $\varepsilon$-dependent part of the integrand and then expand this part in $\varepsilon$ before integrating
\begin{equation} \label{eq: integral proper time with expansion}
    \Delta \tau_{r_\gamma \rightarrow\,r_h}=\int_{0}^{1} \mathrm{d}x\frac{1}{\sqrt{1-x}} \underbrace{\frac{\left(r_\gamma(\varepsilon)-r_h(\varepsilon) \right)}{\sqrt{-\frac{f(x,\varepsilon)}{1-x}\left(1-\frac{h(1,\varepsilon)}{h(x,\varepsilon)} \right)}}}_{U(\varepsilon,x)=\sum_{n=0}^\infty \frac{1}{n!}U'(0,x)\varepsilon^n}\,.
\end{equation}

\subsection{Increased surface gravity}
Static and spherically symmetric, regular black holes suffer from a problem even more catastrophic than the singularity of static vacuum solutions in GR, namely a Cauchy horizon \cite{Hayward:2005gi,Carballo-Rubio:2018pmi, Carballo-Rubio:2019fnb, Carballo-Rubio:2019nel}. The reason underlying this is that static regular black holes necessarily have an inner horizon, because regularity and asymptotic-flatness necessarily lead to two zeros of $g^{rr}(r)$.
Because the region interior to this horizon is, loosely speaking, a region of ``repulsive gravity'' (e.g., in many examples it is a de Sitter core), the inner horizon is a Cauchy horizon.
At a Cauchy horizon, the initial-value problem, set up on a Cauchy hypersurface to the past of the Cauchy horizon, breaks down and the evolution beyond the Cauchy horizon can no longer be continued uniquely. Thus, a Cauchy horizon is not just a single point at which the spacetime is unphysical, but a whole hypersurface. Therefore, static regular black holes suffer from a problem much more catastrophic than a curvature singularity at a single point.

This problem exists in static spacetimes. However, these spacetimes already have a mechanism built in that may prevent such a catastrophic breakdown of predictability. This mechanism is the instability of the inner horizon.\footnote{In the literature, the degree of instability of the Cauchy horizon (exponential vs. power-law growth of perturbations) has been debated \cite{Bonanno:2020fgp,Carballo-Rubio:2021bpr,Carballo-Rubio:2022kad,Franzin:2022wai,Bonanno:2022jjp,Bonanno:2022rvo,McMaken:2023uue,Bonanno:2025bgc}, see \cite{DiFilippo:2022qkl, Carballo-Rubio:2021bpr} for non-perturbative backreaction; recent studies include the dynamical equations for which such black holes are solutions \cite{Frolov:2026rcm,Borissova:2026kwu,DiFilippo:2026jpv,Bueno:2026wjt}. In many studies, the point of view taken has been that the instability of the Cauchy horizon is a problem for the black-hole geometry. Here, we take the opposite  perspective  \cite{ Barcelo:2020mjw, Barcelo:2022gii, Arrechea:2020jtv, DiFilippo:2025kzh}, namely that the instability is a feature, not a bug, because it provides a pathway to obtaining a (necessarily time-dependent) black-hole geometry that does not contain surfaces at which predictability breaks down.} At an inner horizon, fluctuations are strongly blueshifted and accumulate. These destabilize the inner horizon. Supplemented by the appropriate dynamics beyond GR, the inner horizon may evolve outwards, see \cite{Zilberman:2019buh, Hollands:2019whz, Barcelo:2020mjw,  Barcelo:2022gii}. At the same time, Hawking radiation, once included in the theory, destabilizes the outer horizon, such that it evolves inward. This makes it possible that a regular black hole forming from gravitational collapse never develops a true event horizon and an inner Cauchy horizon. Instead, both horizons are trapping horizons, which only exist for a finite lifetime, see \cite{Barenboim:2024dko,Barenboim:2025ckx} for numerical results in a 1+1 dimensional setting that exhibit this mechanism. Recently, this has been studied analytically in \cite{Arrechea:2026rua}. In the 3+1 dimensional setting, this has been studied numerically for a Reissner-Nordström black hole \cite{Boyanov:2025otp}.

We now come to our conjecture pertaining to surface gravity: we require these instabilities to be as effective as possible, in order to prevent a breakdown of predictability. 
Because these instabilities are driven by the surface gravity, this brings us to the conjecture that the surface gravity at the inner horizon should be large. The inner horizon is not necessarily visible to the EFT. Therefore, we formulate a more restrictive conjecture, which ties the surface gravity at the inner and outer horizon together and requires the surface gravity at the outer horizon to be large, 
\begin{align}
     \kappa=\frac{1}{2}\sqrt{h'(r_h)f'(r_h)}&=\frac{1}{2 r_S}\left(\frac{r_S}{r_h}\right)^2 \sqrt{\left( f_1(r_h) - r_h f_1'(r_h) \right) \left( f_1(r_h) - r_h f_2'(r_h) \right)}\\&\overset{!}{>}\kappa_S=\frac{1}{2r_S}
\end{align}
where we have used that $f_2(r_h)=f_1(r_h)$. We impose $\kappa> \kappa_S$ as a nontrivial requirement, because it does not follow from the previous restrictions on $r_h$ and $f_i$ and $f_i'$. In fact, in our expression for $\kappa$, the Schwarzschild value $1/(2 r_S)$ is multiplied by one factor larger than one ($r_S>r_h$) and one factor that can be smaller than one ($f_i<1$ and $f_i'>0$); accordingly, $\kappa$ may be larger or smaller than $\kappa_S$ for metrics that satisfy our previous constraints. 

Let us highlight that of our five conjectures, this one takes a special place for two reasons: first, while the other four conjectures relate to the improved behavior of static regular black holes, this fifth conjecture points beyond the static case and foreshadows behavior we expect to see in time-dependent cases. Second, we consider it more plausible that physically acceptable spacetimes can be constructed that violate this conjecture rather than the other four. This is because a theory may find a way to exhibit physically acceptable time-dependent black-hole spacetimes without going through the mechanism we rely on here, namely an instability in the static spacetime. In addition, we highlight again that drawing a connection between the surface gravity at the inner and outer horizon is a stronger assumption than just that of simplicity.

\subsection{Summary: Tracers of black-hole regularity}
In summary, the five tracers of black-hole regularity are the following:
Our first two tracers refer to the weakening of gravity associated to the regularization of the central curvature singularity. They diagnose this weakening through the behavior of light rays, i.e., null geodesics, in the spacetime and require that surfaces of \emph{strong gravitational effects on light} become more compact, i.e., the horizon and the photon sphere become more compact.\\
Our third and fourth tracers refer to the motion of observers in the spacetime and again encode effects of weakened gravity. One is the extended proper time it takes observers to move between surfaces of given areal radius and which we require to increase. The other is the amount of tidal stretching, or ``spaghettification'' at the horizon, which we require to decrease.\\
We highlight that most of our tracers refer to matter probes of the geometry, either in terms of massless probes (i.e., light-like trajectories) or massive probes (i.e., time-like trajectories). We assume that these follow the corresponding geodesics of the metric in Eq.~\eqref{eq:metric}. This leads us to an important point, namely the validity of field redefinitions, see Sec.~\ref{sec:field_redefinitions}.

Our final tracer refers to a theoretical inconsistency of static, curvature-regular spacetimes, namely that they contain an inner horizon which is a Cauchy horizon. The associated breakdown of the initial-value problem must be prevented, which is possible if the mass-inflation of the Cauchy horizon is as effective as possible. This is realized if the surface gravity at the Cauchy horizon is as large as possible and we conjecture that this is linked to the surface gravity of the outer horizon becoming as large as possible. This tracer is the only one that encodes a notion of stronger gravity, which we consider necessary to avoid a theoretical breakdown even more disastrous than a curvature singularity, namely a Cauchy horizon.\\

We emphasize that the details of these tracers are largely meant as plausible examples, rather than unbreakable conditions on physically acceptable black-hole spacetimes. They serve two purposes: first, they are plausible conditions and it is therefore informative whether or not it is actually possible to satisfy them together early on in the EFT expansion. Second, they serve as examples to highlight the general idea of constraining the EFT from similar considerations. In this spirit, more than one ``island'' of plausible tracers of regularity may exist in the EFT description of black holes.

\subsection{Tracers of black-hole regularity in model spacetimes}

There are model spacetimes of regular black holes, including the Hayward \cite{Hayward:2005gi}, Dymnikova \cite{Dymnikova:1992ux}, Bardeen \cite{Bardeen1968} and Simpson-Visser spacetimes \cite{Simpson:2019mud}.
These are typically postulated without derivation from an action principle, although progress on dynamical principles has been made \cite{Borissova:2026wmn,Eichhorn:2025pgy, Bueno:2024dgm, Bueno:2025zaj}. We stress that, because none of these spacetimes has been derived as a solution in a full theory of quantum gravity, realization or failure of our conjectures for these model spacetimes does not confirm nor invalidate them. Instead, we discuss them in the light of our conjectures in order to assess whether they should be considered useful model spacetimes that may, e.g., guide searches for observational imprints of regularity.

\begin{figure}[!t]
    \centering
    \includegraphics[width=0.495
    \linewidth]{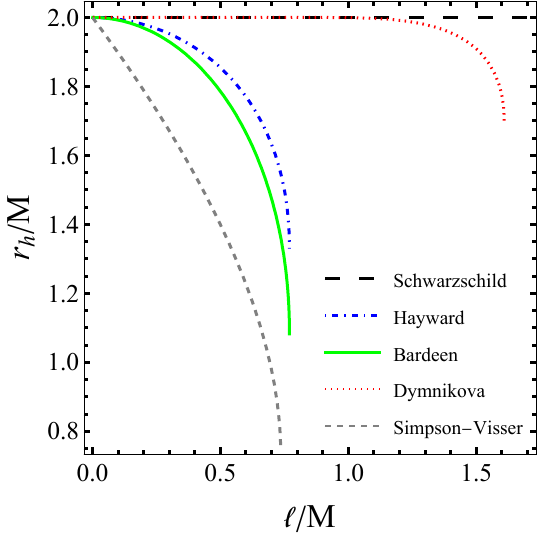}
     \includegraphics[width=0.495
    \linewidth]{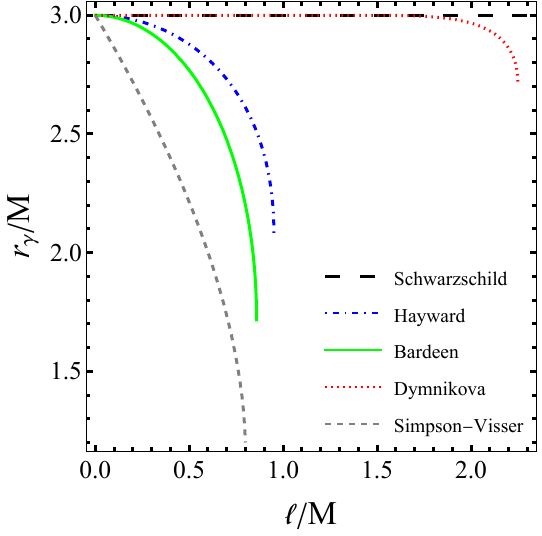}
    \caption{Event horizon $r_h$ and photon sphere $r_\gamma$ locations of regular metrics as functions of the new-physics scale $\ell$. These include the following : Hayward $f(r)=1-\frac{2Mr^2}{2\ell^2M+r^3}$, Bardeen $f(r)=1-\frac{2M}{r}\left(\frac{\ell^2}{r^2}+1 \right)^{-3/2}$, Dymnikova $f(r)=1-\frac{2M}{r}\left(1-e^{-\frac{r^3}{M \ell^2}} \right)$, Simpson-Visser $f(r)=1-\frac{2M}{r}e^{-\frac{\ell}{r}}$. All examples are of the type $f(r)=h(r)$ and fulfill $r_h<2M$, and $r_{\gamma}<3M$.}
    \label{fig: horizon and photon sphere examples}
\end{figure}

The first two tracers, namely increased compactness of the horizon as well as the photon sphere, are realized in these model spacetimes, see Fig.~\ref{fig: horizon and photon sphere examples}. In these cases, an increase in the scale $\ell$ that is associated to regularization of the curvature singularity, strengthens the effects.

\begin{figure}[ht]
    \centering
    \includegraphics[width=0.8\linewidth]{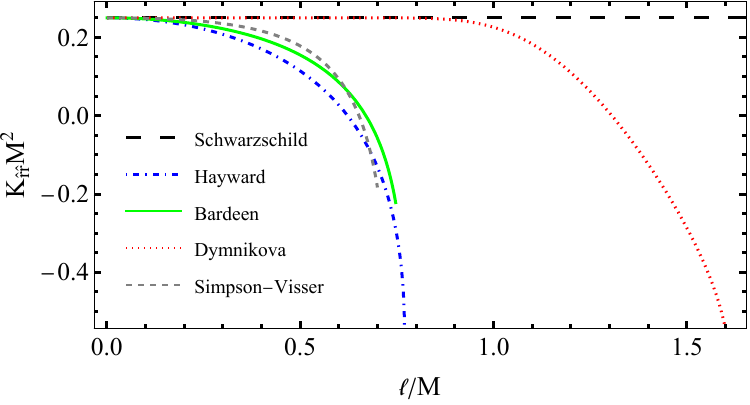}
    \caption{Radial component of tidal-force tensor evaluated at the horizon of regular metrics as a function of the new-physics scale $\ell$ for $M=1$.  These include the following : Hayward $f(r)=1-\frac{2Mr^2}{2\ell^2M+r^3}$, Bardeen $f(r)=1-\frac{2M}{r}\left(\frac{\ell^2}{r^2}+1 \right)^{-3/2}$, Dymnikova $f(r)=1-\frac{2M}{r}\left(1-e^{-\frac{r^3}{M \ell^2}} \right)$, Simpson-Visser $f(r)=1-\frac{2M}{r}e^{-\frac{\ell}{r}}$ All examples are of the type $f(r)=h(r)$.}
    \label{fig: tidal forces at horizon examples}
\end{figure}

Our third and fourth tracer of regularity refers to infalling observers.
Fig.~\ref{fig: tidal forces at horizon examples} shows how the radial component of the tidal-force tensor in the tetrad frame evaluated at the horizon depends on  the new-physics scale $\ell$ for the model spacetimes. In all examples, we find that the tidal forces at the horizon have decreased through the regularization mechanism compared to Schwarzschild, i.e., there is reduced ``spaghettification'' in these spacetimes. 

For our fourth conjecture on an increase in proper time for an observer infalling between photon sphere and horizon, we focus on the generalized Hayward metric \cite{Zhou:2022yio}
\begin{equation}
    f(r)=h(r)=1-\frac{2 M r^{n-1}}{\ell^n+r^n}\,, \quad n\geq 4\quad \text{even}\,.
    \label{eq: generalised hayward}
\end{equation}

\begin{figure}[]
    \centering
    \includegraphics[width=0.6\linewidth]{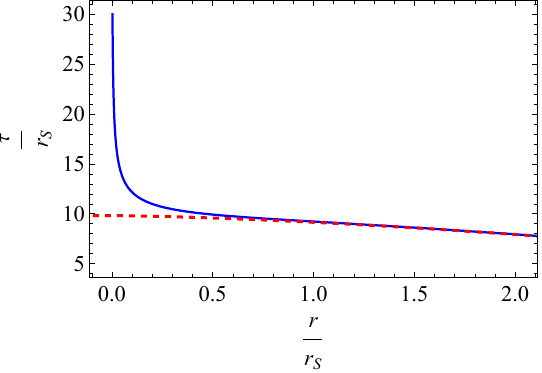}
    \caption{Prolonged proper time in the generalized Hayward metric with respect to the Schwarzschild geometry for $E_{\text{gen. Hayward}}^2=1$. Here, $\ell=0.8$, $R=6r_S$, $n=4$.}
    \label{fig: proper time comparison: Schwarzschild vs. generalised Hayward}
\end{figure}

In Fig.~\ref{fig: proper time comparison: Schwarzschild vs. generalised Hayward}, obtained through numerical integration of Eq.~\eqref{eq: proper time int}, we compare the proper time $\tau$ as a function of $r$ for the Schwarzschild metric and the generalized Hayward metric in the scenario with initial condition $E^2=1$. Note that this corresponds to an observer initially at rest at infinity. We find that \begin{equation}
     \tau^{\text{gen. Hayward}}> \tau_S, 
    \label{eq:delta_tau_comparison}
\end{equation}
for all values $r<R$. Note, that in spherical symmetry, $r$ describes the radius of spheres in the spacetime and therefore has a geometrical significance exceeding that of a mere coordinate function. 

Let us now assume that we have only knowledge of a perturbatively corrected Schwarz-schild metric, which, upon considering all corrections, gives the generalized Hayward metric. 

Expanding Eq.~\eqref{eq: generalised hayward} in powers of the new-physics scale $\ell$, we obtain
\begin{equation}
    f(r)=1-\frac{2M}{r}+\frac{2 M}{r^{n+1}}\ell^n+O(\ell^{2n})\,.
\end{equation}
From now on, we focus on $n=4$. The modified black-hole horizon and photon sphere up to order $l^4$ read
\begin{align}
  r_h^{\text{gen. Hayward trunc.}}=2M-\frac{l^4}{8 M^3}\,, \quad r_{\gamma}^{\text{gen. Hayward trunc.}}=3M-\frac{7 l^4}{81 M^3}\,,
\end{align} 
where more compact black-hole surfaces already appear at a perturbative level.

In the case of the truncated generalized Hayward metric, the proper time then reads
\begin{equation}
    \Delta \tau^{\text{gen. Hayward trunc.}}_{r_\gamma \rightarrow\,r_h}=\sqrt{3}M \left(1+\frac{3}{\sqrt{2}}\,\text{arccot} \left( \sqrt{2}\right) \right)+l^4 \frac{467-180 \sqrt{2} \,\text{arccot}\left(\sqrt{2} \right)}{1620 \sqrt{3}M^3}\,,
\end{equation}
where the numerator of the second term is positive, such that also in this scenario, proper time is prolonged for the truncated generalized Hayward metric.

These model spacetimes exemplify the key challenge of most attempts to construct regular black-hole spacetimes: These attempts remove the curvature singularity, but inevitably (except for the extremal case) introduce a Cauchy horizon into the spacetime. At the same time, these spacetimes contain a hint as to the solution of this problem: the Cauchy horizon is generically unstable, such that a dynamical spacetime may contain trapping horizons as transient features, but be free from Cauchy horizons.

To make this instability as effective as possible, the surface gravity of the inner horizon, which sets the instability timescale, should be large. We expect this to be linked to an increase in the surface gravity of the outer horizon. This expectation is not realized in the example spacetimes, but the simplicity constraint that they satisfy does not restrict the surface gravity directly and so regular black-hole metrics with an increased surface gravity at both horizons should exist.

\section{Against field redefinitions}
\label{sec:field_redefinitions}
We now discuss an important conceptual issue that motivates our selection of terms in the effective action. This issue is the effect of field redefinitions.
Consider the following infinitesimal field redefinition of the metric, 
\begin{equation}
g_{\mu\nu} \rightarrow g_{\mu\nu}'=g_{\mu\nu}+ \delta g_{\mu\nu}= g_{\mu\nu}+ \alpha R_{\mu\nu}+ \beta g_{\mu\nu}R,
\end{equation}
with dimensionful coefficients $\alpha$ and $\beta$. The Einstein-Hilbert action changes by
\begin{eqnarray}
\delta S_{\rm EH}&=&\frac{\delta S_{\rm EH}}{\delta g_{\mu\nu}}\,\delta g_{\mu\nu} = \frac{M_{\rm Pl}^2}{2}\int \mathrm{d}^4x \sqrt{-g}\left(R^{\mu\nu}- \frac{1}{2}g^{\mu\nu}R \right)\,\delta g_{\mu\nu} \nonumber\\
&=& \frac{M_{\rm Pl}^2}{2}\int \mathrm{d}^4x \sqrt{-g}\left(\alpha R_{\mu\nu}R^{\mu\nu} - \frac{\left(2\beta+\alpha\right)}{2}R^2 \right).
\end{eqnarray}
Thus, by a judicious choice of coefficients $\alpha$ and $\beta$, we can remove the quadratic curvature terms $R^2$ and $R_{\mu \nu} R^{\mu \nu}$ in the effective action. Similarly, a subset of higher-order terms proportional to the Ricci tensor can be removed. This would make it seem as if we could simplify our task of analyzing the EFT quite considerably, because we can remove a subset of EFT coefficients in this way. Instead, the field transformation shifts some of the physics encoded in the low-order EFT coefficients into higher-order terms, which can then be dropped, if one works only to low orders in the EFT expansion.

We caution that this procedure results in misleading conclusions drawn about properties of black holes, unless one is very careful. This is because matter couples to the original, untransformed metric $g_{\mu\nu}$, because this is the set of degrees of freedom that we originally obtain by integrating out some heavy degrees of freedom. At a first glance, it may appear irrelevant, which metric matter couples to -- after all, we are analyzing vacuum solutions of the theory. However, most statements about black holes implicitly invoke the presence of matter. Most importantly, this applies even to the event horizon. The causal structure of the metric $g_{\mu\nu}$ and $g_{\mu\nu}'$ is not in general the same, because the field transformation we are considering is not conformal. Thus, determining an event horizon based on the geodesics of $g_{\mu\nu}'$ is only meaningful for metric perturbations, i.e., gravitational waves. Matter is in general non-minimally coupled to $g_{\mu\nu}'$, so that it does not propagate on geodesics of $g_{\mu\nu}'$. Therefore, determining a horizon based on $g_{\mu\nu}'$ is not particularly informative regarding the causal structure of the spacetime that matter will ``see''.
Similarly, the photon sphere of $g_{\mu\nu}'$ does not in general describe the location of marginally unstable circular photon orbits. Because, as highlighted above, most of our tracers of black-hole regularity implicitly or explicitly refer to the trajectories of matter in the spacetime, field redefinitions would in general result in changes of the tracer conditions.

As a consequence, we refrain from any field redefinitions in this work. We also caution that conclusions drawn about black-hole spacetimes after disformal transformations must be regarded with care.

\section{Effective field theory: action and corrections to the Schwarzschild metric}\label{sec: EFT}
We work in the EFT paradigm, in which the Einstein-Hilbert Lagrangian is supplemented by higher-order terms. The higher-order terms have dimensionful couplings, which can be written as an appropriate power of an energy scale $\Lambda_{\rm UV}$ times a dimensionless coefficient.
We make a naturalness assumption, i.e., we assume that the suppression scale $\Lambda_{\rm UV}$ is the same for all higher-order terms and the dimensionless coefficients are numbers of $\mathcal{O}(1)$.\footnote{We caution that such an assumption may be ``unnatural'' in a given UV completion; for instance, asymptotically safe gravity may change the scaling dimension of couplings such that distinct effective suppression scales emerge, see \cite{Brenner:2024bps} for an explicit example within the Standard Model effective field theory.} The general form of the action, in a derivative expansion, is therefore given as
\begin{equation}
S = \frac{M_{\rm Pl}^2}{2}\int \mathrm{d}^4x \sqrt{-g} \left(R + \frac{1}{\Lambda_{\rm UV}^2} \mathcal{L}_4 + \frac{1}{\Lambda_{\rm UV}^4} \mathcal{L}_6 + \frac{1}{\Lambda_{\rm UV}^6} \mathcal{L}_8+ \mathcal{O} \left( \frac{1}{\Lambda_{\rm UV}^8} \right)\right)\label{eq:action}\,,
\end{equation}
where the $\mathcal{L}_n$ are composed of operators with $n$ derivatives.
We do not consider a cosmological constant term, because its effects are irrelevant for the properties of black-hole spacetimes that we are interested in here, for realistic values of the cosmological constant.
The higher-order terms are treated at a perturbative level. Therefore, we can make an ansatz for a spherically symmetric, static vacuum solution of the form

\begin{equation}
    \begin{aligned} 
        \mathrm{d} s^2 =&- \left( 1-\frac{r_S}{r} + \varepsilon^4 b(r)+\varepsilon^6 l(r) + \mathcal{O}\left(\varepsilon^8\right) \right) \mathrm{d} t^2 \\&+ \left( 1-\frac{r_S}{r} + \varepsilon^4 a(r) +\varepsilon^6 k(r) + \mathcal{O}\left(\varepsilon^8\right)\right)^{-1}\, \mathrm{d} r^2 + r^2 \mathrm{d}\Omega^2,
    \end{aligned}
    \label{eq:metric_corrected}
\end{equation}
where
\begin{equation}
    \varepsilon = \frac{1}{r_S\Lambda_{\rm UV}}\ll 1\,,
\end{equation}
is the perturbative dimensionless expansion parameter. The typical length scale of new physics, $\Lambda_{\rm UV}^{-1}$, is therefore much smaller than the Schwarzschild radius for the black holes that we are interested in here.
There is no term $\mathcal{O}(\varepsilon^2)$, because the leading-order perturbative contributions from $\mathcal{L}_4$ vanish on a Schwarzschild background --  or more generally, a Ricci-flat background -- and the first contribution from $\mathcal{L}_4$ instead occurs at $\mathcal{O}(\varepsilon^8)$.  $\mathcal{L}_6$ gives rise to the terms $\mathcal{O}(\epsilon^4)$ and $\mathcal{L}_8$ to the terms $\mathcal{O}(\epsilon^6)$. Mixing of different EFT orders can for the first time occur at $\mathcal{O}(\epsilon^{8})$, which we do not include in our analysis.\newline

We now construct the terms $\mathcal{L}_{4,6,8}$ explicitly. To do so, we use that on a static and spherically symmetric background, all parity-odd operators vanish identically and therefore do not contribute to corrections to the Schwarzschild metric at any order. Consequently, we neglect such operators.  
 
In detail, at order four in derivatives the most general Lagrangian is
\begin{equation}
    \mathcal{L}_4 = a_1 R^2 + a_2 R_{\mu \nu} R^{\mu \nu} + a_3 \mathcal{G}\,,
    \label{eq: quartic correction}
\end{equation}
where the Gauss-Bonnet term
\begin{equation}
    \mathcal{G}=R^2-4R^{\mu \nu}R_{\mu \nu}+R^{\alpha \beta \gamma \delta}R_{\alpha \beta \gamma \delta},
\end{equation}
is a topological invariant in four dimensions and thus cannot contribute to the equations of motion. Thus, we can neglect it and consider only the operators $R^2$ and $R^{\mu \nu}R_{\mu \nu}$.
For our choice of Ricci-flat background metric, none of the terms in Eq.~\eqref{eq: quartic correction} contribute to corrections at leading order, because the terms are quadratic in the Ricci tensor. Thus, we can set $a_1=0=a_2$ without loss of generality, up to this order in the derivative expansion.

The most general non-degenerate parity-invariant Lagrangian density formed with curvature invariants with six derivatives is \cite{Fulling:1992vm,Cano:2019ore,Bueno:2019ltp}
\begin{equation}
\begin{aligned} \label{eq: action order six}
\mathcal{L}_6&=b_1 R_{\mu\nu }^{\ \ \rho\sigma }R_{\rho\sigma }^{\ \ \delta\gamma }R_{\delta\gamma }^{\ \ \mu\nu }+b_2 R_{\mu\nu\rho\sigma }R^{\mu\nu\rho\sigma }R+b_3 R_{\mu\nu\rho\sigma }R^{\mu\rho}R^{\nu\sigma}  \\&\,\,\,\,\,\,\,+b_4 R_{\mu}^{\ \nu}R_{\nu}^{\ \rho}R_{\rho}^{\ \mu}+b_5 R_{\mu\nu }R^{\mu\nu }R + b_6 R^3 + b_7 \nabla_{\sigma}R_{\mu\nu} \nabla^{\sigma}R^{\mu\nu}+b_8 \nabla_{\mu}R\nabla^{\mu}R\,\\
&=b_1 R_{\mu\nu }^{\ \ \rho\sigma }R_{\rho\sigma }^{\ \ \delta\gamma }R_{\delta\gamma }^{\ \ \mu\nu }+b_2 R_{\mu\nu\rho\sigma }R^{\mu\nu\rho\sigma }R + \text{Ricci-squared terms}.
\end{aligned}
\end{equation}
For our purposes, because the background metric is Ricci-flat, we only need to consider the $b_1$ and $b_2$ terms. 

The parity-invariant terms at order eight in derivatives that do not vanish under the Ricci-flatness condition are  
\begin{equation}
    \begin{aligned}\label{eq: action order eight Riem basis}
        \mathcal{L}_8
        &=c_1 \left(R_{\alpha \beta \gamma \delta} R^{\alpha \beta \gamma \delta}\right)^2+c_2 {R}_{\quad \mu \nu}^{\kappa \lambda} R^{\mu \nu}_{\quad \rho \sigma}R^{\rho \sigma}_{\quad \tau \omega}R^{\tau \omega}_{\quad \kappa \lambda}+c_3\, R \, R^{\mu \nu}_{\quad \rho \sigma}R^{\rho \sigma}_{\quad \tau \omega}R^{\tau \omega}_{\quad \mu \nu}\\&\,\,\,\,\,\,\,+c_4 R_{\mu \nu \rho \sigma} R^{\mu \nu \rho \sigma}\, \Box\, R+c_5 R^{\mu \nu \alpha \beta}R^{\rho \sigma}_{\quad \alpha \beta}\nabla_{\mu}\nabla_{\rho} R_{\nu \sigma}\,.
    \end{aligned}
\end{equation}
A derivation of the relevant basis elements based on
\cite{Fulling:1992vm} can be found in Appendix~\ref{app:constructing_the_eft}.  
We do not exclude terms that can be removed by field redefinitions, such as the $b_2,c_3,c_4$ and $c_5$  terms. This is critical, because field redefinitions result in a modified coupling to matter, see Sec.~\ref{sec:field_redefinitions}. Therefore, quantities such as the horizon location can only be computed correctly, if the modified coupling to matter -- in the case of the horizon massless fields -- is properly accounted for.
In fact, of the quantities we consider, only the surface gravity of a black hole is an observable invariant under field redefinitions  in which the modified coupling to matter is neglected~\cite{Jacobson:1993vj}. All other observables require a careful treatment that involves the modified -- and typically non-minimal -- coupling to matter. Because it is much simpler to work with minimally coupled matter, we simply refrain from performing any field redefinitions and work with the assumption of matter that is minimally coupled to that metric.

Next, we determine the functions $a(r),\, b(r),\, k(r)$ and $l(r)$ in Eq.~\eqref{eq:metric_corrected} in terms of the couplings of the EFT by solving the equations of motion.
We obtain terms $\mathcal{O}(\varepsilon^{n})$ by inserting the $\mathcal{O}(\varepsilon^{n})$ terms in the metric ansatz into the Einstein-Hilbert part of the action. These have to match what we obtain by inserting the Schwarzschild part of the action into the $\mathcal{O}(\Lambda_{\rm UV}^{-n})$ part of the Lagrangian.

For example, the equations of motion at order six read
\begin{align}
        0&=\frac{r^8 a(r)-r^7 r_Sb(r)+(r-r_S)\left(6 r_S^6 \left[b_1\left(-9r+5r_S\right)+6b_2 (-4 r +3r_S) \right]+r^8 b'(r)\right)}{r^6 (r-r_S)^2}\,,\\
        0&=r\frac{a(r)+ra'(r)}{r_S-r}+\frac{6 r_S^6 \left(b_1\left(45 r-49r_S\right)+6b_2 \left( 10r-11r_S\right)\right)}{r^6 (r_S-r)}\,,
    \end{align}
where primes denote derivatives with respect to $r$.
They are solved by
\begin{align}
 b(r)&=\frac{C_1}{r_S}+\left( -1+\frac{r_S}{r}\right)C_2+\frac{r_S^6}{r^6} \left[ -12 b_2 + \left(5b_1+18b_2\right)\frac{r_S}{r} \right]\,, \\
    a(r)&=\frac{C_1}{r}+\frac{r_S^6}{r^6} \left[ \left(54 b_1 + 72 b_2\right) - \left(49 b_1 + 66 b_2\right) \frac{r_S}{r} \right]\,,
\end{align}
where $C_{1,2}$ are integration constants.
By demanding asymptotic flatness and that $M$ still represents the ADM mass of the black hole, we obtain
\begin{equation}
    C_1=C_2=0\,.
\end{equation}
At order eight, we obtain similar equations and can set the integration constants to zero as well.
In the end, the metric perturbations that are related to the higher-derivative corrections are given by
\begin{equation}
\begin{aligned}\label{eq: metric function results}
        b(r)&=\frac{r_S^6}{r^6} \left[ -12 b_2 + \left(5b_1+18b_2\right)\frac{r_S}{r} \right]\,, \\
        a(r)&=  \frac{r_S^6}{r^6} \left[ \left(54 b_1 + 72 b_2\right) - \left(49 b_1 + 66 b_2\right) \frac{r_S}{r} \right]\,,\\
         l(r)&=\frac{r_S^8}{4 r^{10}}\left[-1440 c_4 r^2+2 r r_S\left(8 \left(32 c_1+6 c_2-3 c_3+243 c_4\right)+27 c_5\right)\right. \\& \left. \hspace{1.5 cm}-r_S^2 \left(8 \left(44 c_1+5 c_2-9 c_3+324 c_4\right)+63 c_5\right)\right]\,, \\
        k(r)&=\frac{r_S^8}{4 r^{10}}\left[720 r^2 \left(16 c_4+c_5\right)+18 r r_S\left(128 c_1+24 \left(2 c_2+c_3-61 c_4\right)-91 c_5\right)\right. \\& \left. \hspace{1.5 cm}+r_S^2 \left(909 c_5-8 \left(268 c_1+101 c_2+51 \left(c_3-36 c_4\right)\right)\right) \right]\,.
    \end{aligned}
\end{equation}
Due to spherically symmetry and staticity, the coupling space at order eight in derivatives is actually not 5-dimensional but only 4-dimensional. We can make this visible by defining 
\begin{equation}
    z_1=c_1-\frac{3}{8}c_3,\qquad  z_2=c_2+\frac{3}{2} c_3\,.
\end{equation}
The metric corrections at order eight in derivatives then read
\begin{equation}
\begin{aligned}
    l(r)&=\frac{r_S^8}{4 r^{10}} \left[-144 c_5 \left(10 r^2-27 r r_S+18 r_S^2\right)+9 c_5 r_S(6 r-7 r_S) \right. \\ & \left. \hspace{1.5 cm}+8 r_S(64 r z_1+12 r z_2-44 r_Sz_1-5 r_Sz_2)\right]\,, \\
        k(r)&=\frac{r_S^8}{4 r^{10}}\left[144 c_4 \left(80 r^2-183 r r_S+102 r_S^2\right)+9 c_5 \left(80 r^2-182 r r_S+101 r_S^2\right)\right. \\& \left. \hspace{1.5cm} +8 r_S(288 r z_1+108 r z_2-268 r_Sz_1-101 r_Sz_2)\right]\,,
\end{aligned}
\end{equation}
such that we only have four free parameters, namely $z_1$, $z_2$, $c_4$ and $c_5$.
\section{How the effective field theory ``sees'' the onset of black-hole regularity} \label{sec: EFT results}
In Sec.~\ref{sec:conjectures}, we have introduced a set of properties that one may expect regular black holes in quantum gravity to have. We have also made the assumption that these properties are already visible in the EFT, i.e., we expect various manifestations of regularity to build up gradually, order by order in the EFT expansion. This motivates the conjecture that the five properties in Sec.~\ref{sec:conjectures} single out a preferred range of the couplings at each order in the EFT. In other words, the conjecture that each of the properties in Sec.~\ref{sec:conjectures} holds, can be viewed as a type of swampland conjecture. It distinguishes those EFTs that give rise to regular black holes with particular properties in their UV completion and do so order by order, from those EFTs that do not.
The signs of regularity strengthen continuously with each EFT order, if we demand that for $n,m \in \{2,4,6,8,...\}\,, \quad m<n$
\begin{eqnarray} 
r_h<r_S&:=&r_h^{(2)}\,, \quad\quad \quad \quad r_h^{(n)}<r_h^{(m)}\,,\label{eq_ EFT constraints1} \\
    r_\gamma<3 M&:=&r_\gamma^{(2)}\,, \quad\quad\quad  \quad  r_\gamma^{(n)}<r_\gamma^{(m)}\,,\\
    K_{\Hat{r}\Hat{r}}(r_h)<
   \frac{1}{r_S^2}&:=&\left(K_{\Hat{r}\Hat{r}}(r_h)\right)^{(2)}\,, \quad \left(K_{\Hat{r}\Hat{r}}(r_h)\right)^{(n)}<\left(K_{\Hat{r}\Hat{r}}(r_h)\right)^{(m)}\,,\\ 
\Delta \tau>\Delta\tau_s&:=&\Delta\tau^{(2)}\,, \quad\quad\quad \quad \Delta \tau^{(n)}>\Delta \tau^{(m)}\,, \\\kappa >
     \kappa_s&:=&\kappa^{(2)}=\frac{1}{2r_S}\,, \quad  \kappa^{(n)}> \kappa^{(m)}\,.\label{eq_ EFT constraints5}
\end{eqnarray}

\begin{figure}[!t] 
  \begin{minipage}[l]{0.4\linewidth}
    \includegraphics[width=\linewidth]{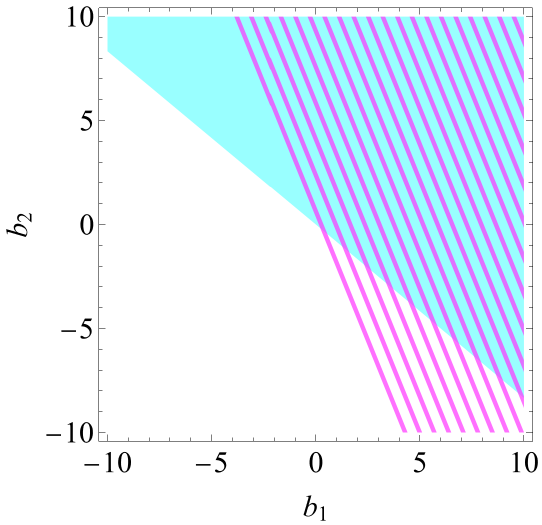}
\includegraphics[width=\linewidth]{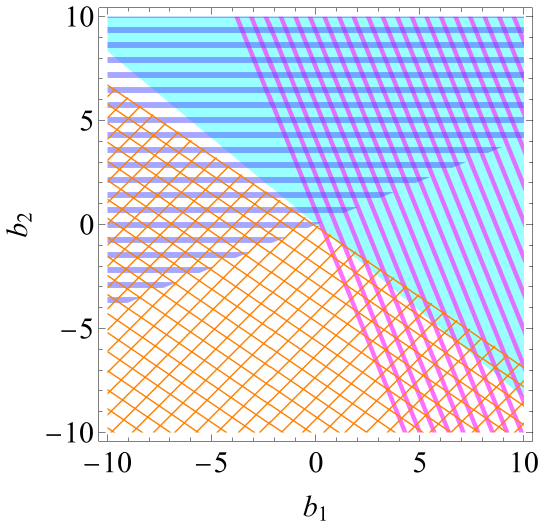}
  \end{minipage} 
  \begin{minipage}[l]{0.55\linewidth}
   \includegraphics[width=0.735\linewidth]{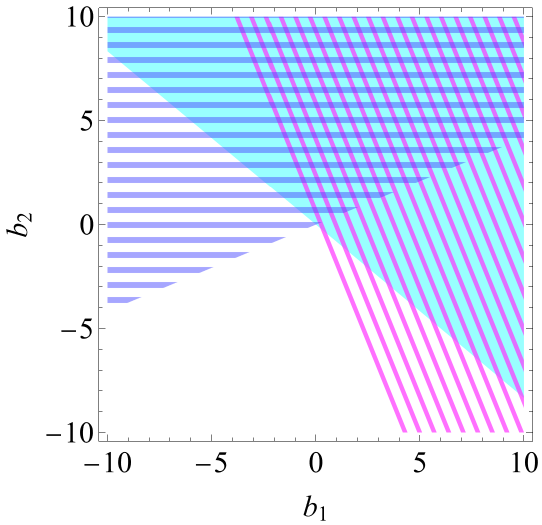}
   \includegraphics[width=\linewidth]{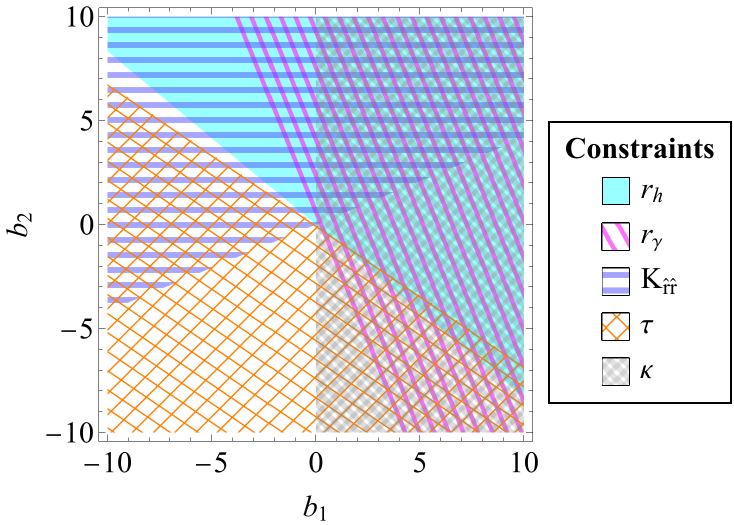}
  \end{minipage} 

\caption{Constraint surfaces of order six couplings $b_1,b_2$. Maximally four constraints can be fulfilled simultaneously. }
    \label{fig: couplings 6}
\end{figure}

The horizon of the EFT corrected spacetime \eqref{eq:metric_corrected} is located at
\begin{equation}
\begin{aligned}\label{eq: eft horizon}
    r_h
    &=r_S\left( 1 - \varepsilon^4 \left[ 5b_1 + 6b_2 \right]+  \varepsilon^6 \frac{1}{4}(144 c_4+9c_5-8(20z_1+7z_2))\right)\,.
    \end{aligned}
\end{equation}
The radius of the photon sphere reads
\begin{equation}
\begin{aligned}\label{eq: eft photon sphere}
    r_{\gamma}
    &=3 M-\varepsilon^4\frac{64}{81} M  (5 b_1+2 b_2)-\varepsilon^6\frac{128  (216 c_4+45 c_5+1408 z_1+368 z_2 )}{6561}M\,.
    \end{aligned}
\end{equation}
The radial tidal tensor component in the tetrad of an observer moving on a radial geodesic evaluated at the horizon is given by 
\begin{equation}
    K_{\hat{r} \hat{r}}(r_h)=\frac{1}{r_S^2}+\frac{2 (5 b_1-12 b_2)}{r_S^2}\varepsilon^4+\frac{\left(1728c_4+9 c_5+2624 z_1+544 z_2\right)}{4 r_S^2}\varepsilon ^6\,.
\end{equation}
The proper time interval required for a radially infalling time-like observer starting from rest at the photon sphere \eqref{eq: eft photon sphere} and then proceeding to freely fall, to reach the horizon \eqref{eq: eft horizon} is obtained by the integral \eqref{eq: integral proper time with expansion}. The result is 
\begin{equation}
\Delta\tau^{\text{EFT}}_{r_\gamma \rightarrow r_h}=\Delta\tau^S_{r_\gamma \rightarrow r_h}+\varepsilon^4 \Delta\tau_{(6)}^{\text{corr}}+\varepsilon^6 \Delta\tau_{(8)}^{\text{corr}}\,,
\end{equation}
with the time interval in the unperturbed Schwarzschild spacetime
\begin{align} \label{eq: proper time interval schwarzschild}
\Delta\tau^S_{r_\gamma \rightarrow r_h}&=M \left(\sqrt{3}+3 \sqrt{\frac{3}{2}} \text{arccot} \left(\sqrt{2}\right)\right)\,,
\end{align} 
and the EFT corrections at order six
\begin{equation}
\begin{aligned} \label{eq: proper time corr 6}
    \Delta\tau_{(6)}^{\text{corr}}=\frac{8 M}{25515 \sqrt{3}} &\left( b_1\, \left( 9450   \sqrt{2}\pi-39707-37800 \sqrt{2} \, \text{arcccot}\left(\sqrt{2-\sqrt{3}}\right)\right) \right. \\ &\left. +b_2 \, \left(5670 \sqrt{2} \left(\pi -4 \, \text{arcccot}\left(\sqrt{2-\sqrt{3}}\right)\right)-72888 \right) \right)\,.
\end{aligned}
\end{equation}
At order eight, we get 
\begin{equation}
\begin{aligned}
    \Delta\tau_{(8)}^{\text{corr}}=\frac{4 M}{98513415 \sqrt{3}} &\left(224\, z_1  \left(-4787640 \sqrt{2} \, \arccos\left(\sqrt{\frac{2}{3}}\right)+5068573\right) \right. \\ &\left. +56\, z_2  \left(603839-5096520 \sqrt{2} \arccos \left(\sqrt{\frac{2}{3}}\right)\right)\right. \\ &\left.+ 1728\, c_4  \left(333647-135135 \sqrt{2} \, \text{arccos}\left(\sqrt{\frac{2}{3}}\right)\right)\right. \\ &\left.-9 \,c_5  \left(3052913+3243240 \sqrt{2} \, \text{arccos}\left(\sqrt{\frac{2}{3}}\right)\right)\right) \,.
\end{aligned}
\end{equation}
Finally, the surface gravity reads
\begin{equation}
    \kappa = 
    \frac{1}{2r_S} \left( 1 + \varepsilon^4 \left[2b_1\right] +8 \varepsilon^6 \left[ 4z_1+z_2\right]\right)\,.
\end{equation} 
Because $\kappa$ is
invariant under field redefinitions, it only depends
on $b_1$, $c_1$ and $c_2$, the couplings of the operators that cannot be removed by field redefinitions.

Tab.~\ref{tab:fneqh} shows the constraints on the EFT couplings imposed by Eqs.~\eqref{eq_ EFT constraints1}-\eqref{eq_ EFT constraints5} at order six and order eight in derivatives, respectively. 
In the order six case we can, at most, fulfill four different constraints at once. This is probably not too surprising, given that we are constraining only two couplings. It is, therefore, already nontrivial that there is a non-empty overlap of four out of five constraints and that we can even choose different sets of four constraints to fulfill simultaneously. This is visualized in Fig.~\ref{fig: couplings 6}, where we display all five constraints in order to see all possible overlaps of the constraint regions.

\begin{figure}[!t]
\centering
    \includegraphics[width=0.4\linewidth]{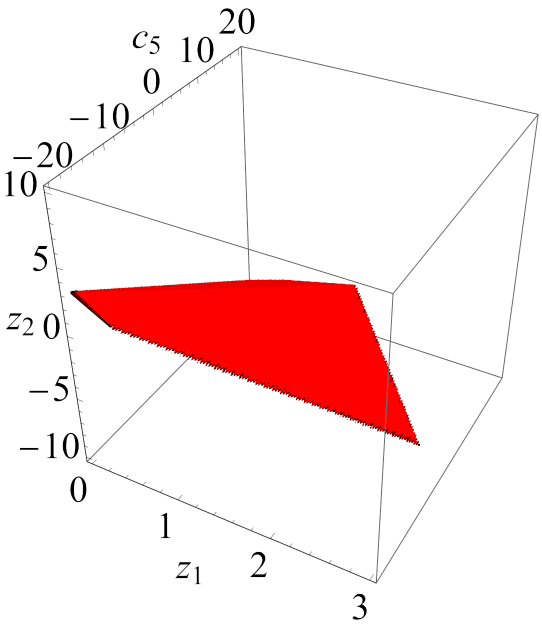}    
    \caption{Polygon of combined constraints for order eight couplings. We show a foliation of the four dimensional coupling space for $c_4=-1$. Other negative values of $c_4$ simply lead to an overall rescaling, because they correspond to a multiplication of the full order-8 EFT action by a positive constant. }
    \label{fig: order eight all constraints}
\end{figure}

However, in the order eight case we can fulfill all five constraints simultaneously. Fig.~\ref{fig: order eight all constraints} shows only the polygon-shaped region, in which all five constraints are met simultaneously. This is displayed via a 3-dimensional slice of the 4-dimensional parameter space for one of the couplings set to a fixed value, without loss of generality.

\begin{table}[hbpt]
\label{tab:fneqh}
\begin{tabular}{ |p{2.85cm}|p{3.05cm}|p{3.15cm}| p{2.0cm}|p{2.0cm}|}
 \hline
 $r_h$ constraint & $r_\gamma$ constr. &$K_{\Hat{r}\Hat{r}}(r_h)\,\,$ constr. &$\tau$ constr. &$\kappa$ constr. \\
 \hline
 \hline
$5b_1+6b_2>0$   & $5b_1+2b_2>0$ &$5b_1-12b_2<0$ &$ \Delta\tau_{(6)}^{\text{corr}}>0$&   $b_1>0$\\
\hline
$144 c_4+9c_5-8(20z_1+7z_2)<0$&   $216c_4+45c_5+1408z_1+368z_2 >0$ &$1728 c_4+9 c_5+2624 z_1+544 z_2<0$& $ \Delta\tau_{(8)}^{\text{corr}}>0$&$4z_1+z_2>0$\\
 \hline
\end{tabular}
\caption{Constraints on EFT couplings from observables.}
\end{table}

\section{Effective field theory with additional light fields} \label{sec: EFT with scalar field}

In the previous chapter, we found that in the pure-gravity sector of the EFT, it is not possible to fulfill all conjectures of black-hole regularity at order six in derivatives. The pure-gravity EFT implicitly assumes that no additional light degrees of freedom mediate gravity. This is not always the case. For instance, string theory  
 may also give rise to light scalar degrees of freedom in the gravitational sector
 \cite{Scherk_1974, Boulware_1986,Gross:1986mw, Gross:1986iv,Kanti:1995vq, Campbell:1990fu, Cano:2021rey, Cano:2019ore}.
 From a different perspective, modified gravity settings that are explored in cosmology often contain light scalars, e.g., to model dynamical dark energy \cite{Wetterich_1988, Ratra_1988, Caldwell:1997ii,Doran:2000jt, Krishnan:2021dyb}.

\begin{figure}[ht] 
  \begin{minipage}[l]{0.4\linewidth}
    \includegraphics[width=\linewidth]{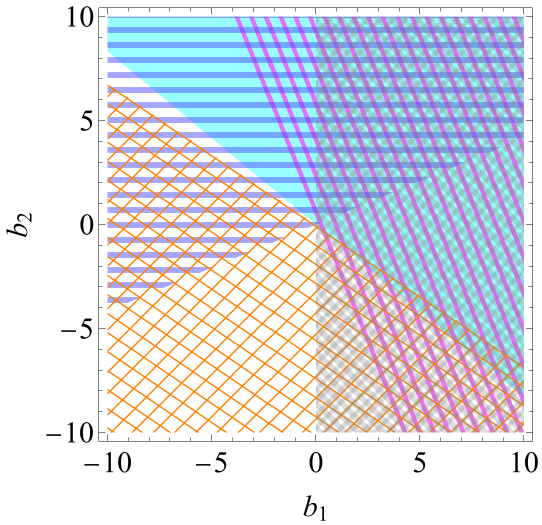}
\includegraphics[width=\linewidth]{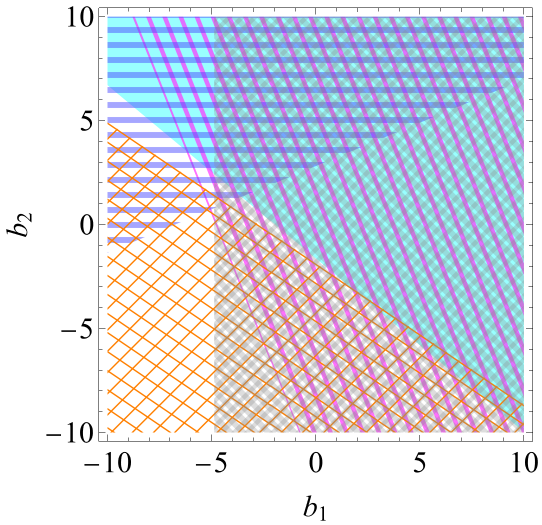}
  \end{minipage} 
  \begin{minipage}[l]{0.55\linewidth}
   \includegraphics[width=0.735\linewidth]{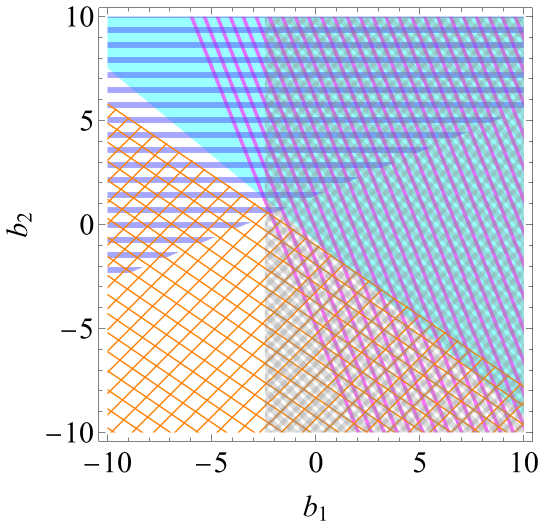}
   \includegraphics[width=\linewidth]{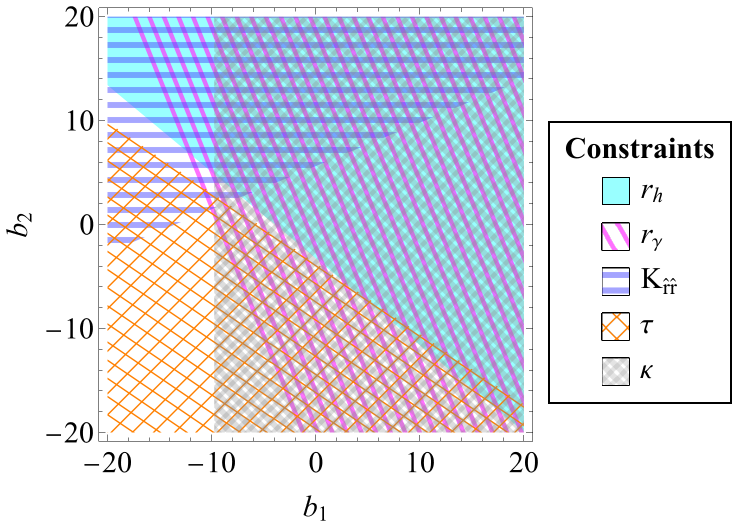}
  \end{minipage} 

\caption{Constraint surfaces at order six in derivatives with scalar hair. We show the constraint regions in the $b_1-b_2$ plane with increasing values of $b_3$. From top left to bottom right: $b_3^2=0, \, 0.5, \, 1, \, 5$. Maximally four constraints can be fulfilled simultaneously. }
    \label{fig: couplings 6 with scalar hair}
\end{figure}

Against this background, we now extend our analysis. We consider the only
non-minimal coupling of gravity to a massless scalar field which is non-trivial on Ricci-flat backgrounds and to this order in the EFT. The corresponding scalar-field action reads
\begin{equation}
\begin{aligned}
S_\phi &= \frac{M_{ Pl}^2}{2}\int \mathrm{d}^4x \sqrt{-g} \left(-\frac{1}{2}\left( \partial_{\mu} \phi \partial^{\mu} \phi\right)+\frac{1}{\Lambda_{\rm UV}^2}b_3 \,\phi\, R_{\alpha \beta \gamma \delta} R^{\alpha \beta \gamma \delta}+ \mathcal{O} \left( \frac{1}{\Lambda_{\rm UV}^6} \right)\right)\label{eq:action phi}\,.
\end{aligned}
\end{equation}
The equation of motion of the scalar field is
\begin{equation}
    \Box \phi=-\frac{b_3}{\Lambda_{\rm UV}^2}R_{\alpha \beta \gamma \delta} R^{\alpha \beta \gamma \delta}\,.
\end{equation}
This implies
that the scalar must be of order $\varepsilon^2$, meaning, it  
only contributes to EFT corrections at order $\varepsilon^4$. 
To see the effects of introducing the scalar field, we only need to consider
\begin{equation}
\begin{aligned}
S_\text{tot} = \frac{M_{ Pl}^2}{2}\int \mathrm{d}^4x \sqrt{-g} \Bigg(&R-\frac{1}{2}\left( \partial_{\mu} \phi \partial^{\mu} \phi\right)+\frac{1}{\Lambda_{\rm UV}^2}b_3 \phi R_{\alpha \beta \gamma \delta} R^{\alpha \beta \gamma \delta}\\&+\frac{1}{\Lambda_{\rm UV}^4} \left(b_1 R_{\mu\nu }^{\ \ \rho\sigma }R_{\rho\sigma }^{\ \ \delta\gamma }R_{\delta\gamma }^{\ \ \mu\nu }+b_2 R_{\mu\nu\rho\sigma }R^{\mu\nu\rho\sigma }R \right)+ \mathcal{O} \left( \frac{1}{\Lambda_{\rm UV}^6} \right)\Bigg)\label{eq:action phi}\,.
\end{aligned}
\end{equation}
For our spherically symmetric static metric ansatz \eqref{eq:metric_corrected}, the equation of motion of the scalar field to leading order is solved by
\begin{equation}
    \phi=\frac{8 b_3 M \left(4 M^2+3 M r+3 r^2\right)}{ 3 r^3}\varepsilon^2\,,
\end{equation}
where we imposed regularity at $r= r_h$ and a vanishing scalar field at infinity. Solving the equations of motion for $f(r)$ and $h(r)$ while imposing asymptotic flatness and requiring that $M$ represents the ADM mass, we find
\begin{equation}
\begin{aligned}
    f(r)&=1-\frac{r_S}{r}+\varepsilon ^4\frac{r_S^6}{r^6} \left[(54b_1+72b_2)-(49b_1+66b_2)\frac{r_S}{r} \right]\\ & \quad \quad+
   \varepsilon ^4 \frac{r_S^2}{15 r^7} b_3^2 \left(60 r^5+30 r^4 r_S+260 r^3 r_S^2+15 r^2 r_S^3+12 r r_S^4-230 r_S^5\right)\,,\\
    \vspace{1cm}
    h(r)&=1-\frac{r_S}{r}+\varepsilon ^4 \frac{r_S^6}{r^6}\left[-12 b_2+(5b_1+18 b_2)\frac{r_S}{r} \right]\\& 
    \quad \quad+\varepsilon ^4 \frac{r_S^3}{15 r^7} b_3^2 \left( 10 r^4+130 r^3 r_S+33 r^2 r_S^2+24 r r_S^3-50 r_S^4\right)\,.
       \end{aligned} 
    \end{equation}
Next, we consider the observables that constitute tracers of regularity. The horizon is located at 
\begin{equation}
    r_h=r_S\left(1-\varepsilon^4 \left(5 b_1+6b_2 + \frac{49}{5}b_3^2\right)\right)\,,
\end{equation}
while the radius of the photon sphere is
\begin{equation}
 r_{\gamma}=3M-\varepsilon^4 \frac{64 M}{81}\left(5b_1+2 b_2+\frac{961}{40} b_3^2\right)\,.
\end{equation}
The radial component of the tidal tensor in the tetrad frame of an observer moving on a radial geodesic evaluated at the horizon becomes
\begin{equation}
    K_{\Hat{r}\Hat{r}}(r_h)=\frac{1}{r_S^2}+\frac{2 \left(25 b_1-60 b_2+169 b_3^2\right)}{5 \,r_S^{2}}\varepsilon ^4 \,.
\end{equation}
The proper time interval of a radial time-like geodesic starting from rest at the photon sphere and reaching the event horizon is
\begin{equation}
\Delta\tau^{\phi-\text{EFT}}_{r_\gamma \rightarrow r_h}=\Delta\tau^S_{r_\gamma \rightarrow r_h}+\varepsilon^4 \Delta\tau_{(6)}^{\text{corr}}+\varepsilon^4 \Delta\tau_{(\phi)}^{\text{corr}}\,,
\end{equation}
where $\Delta\tau^S_{r_\gamma \rightarrow r_h}$ is defined in \eqref{eq: proper time interval schwarzschild} and $\Delta\tau_{(6)}^{\text{corr}}$ in \eqref{eq: proper time corr 6}. The correction coming from the scalar field reads 
\begin{equation}
    \Delta\tau_{(\phi)}^{\text{corr}}=\frac{8  M b_3^2\left(-1434061+2009070 \sqrt{2}\, \text{arctan}\left(\sqrt{2}-\sqrt{3}\right)\right)}{382725 \sqrt{3}}\,.
\end{equation}
Finally, the surface gravity is
\begin{equation}
     \kappa 
     =\frac{1}{2 r_S}\left(1+2 b_1\varepsilon^4+ \frac{146}{15}b_3^2\varepsilon^4\right)\,.
\end{equation}
Tab.~\ref{tab:scalar-order6} and Fig.~\ref{fig: couplings 6 with scalar hair} summarize the constraints arising from all observables considered. In the extended parameter space that includes the $b_3^2$ coupling, no  region exists in which all five constraints can be met simultaneously.

\begin{table}[hbpt]
\label{tab:scalar-order6}
\begin{tabular}{ |p{2.5cm}|p{2.5cm}|p{3.0cm}| p{2.3cm}|p{2.5cm}| }
 \hline
 $r_h$ constraint& $r_\gamma$ constraint&$ K_{\Hat{r}\Hat{r}}(r_h)\,\,$constraint&$\tau$ constraint&$\kappa$ constraint \\
 \hline
 \hline
$5b_1+6b_2+\frac{49}{5}b_3^2>0$   & $5b_1+2 b_2+\frac{961}{40} b_3^2>0$     &$25 b_1-60 b_2+169 b_3^2<0$&$ \Delta\tau_{(\phi)}^{\text{corr}}+ \Delta\tau_{(6)}^{\text{corr}}>0$&   $15 b_1+73b_3^2>0$\\
 \hline
\end{tabular}
\caption{Constraints on scalar-tensor EFT couplings from observables.}
\end{table}

\section{String theory and asymptotic safety as testing grounds} \label{sec: comparison to ST and ASQG}
Until now, we have worked purely within the EFT of gravity. However, there are concrete examples of frameworks for the UV properties of gravity. In two such frameworks, namely string theory and asymptotically safe gravity, coefficients of the effective action are known \cite{Kawai:1985xq,Cano:2019ore,Cano:2021rey, Bern:2021ppb, Melville:2024zjq,Baldazzi:2023pep}. In addition, both frameworks are expected to give rise to singularity-free geometries, with arguments reviewed in \cite{Eichhorn:2022bgu,Platania:2023srt,Nicolini:2023hub}. Thus, they constitute testing grounds for our tracer conditions.

\subsection{Effective field theory with field redefinitions}
String-theoretic EFTs are usually derived in a frame in which there are no Ricci tensors. Thus, the EFTs in  \cite{Cano:2019ore, Cano:2021rey} cannot directly be compared to our tracer conditions. Instead, we perform a disformal transformation on the metric and thereby map our action into a transformed action that can directly be compared to string-theoretic EFTs.  We highlight that we do \emph{not} use the disformal frame to derive the implications of the tracer conditions. Instead, the tracer conditions remain those formulated in the original frame. We only map the resulting constraints on the couplings to the couplings in the EFT without Ricci tensors. Deriving the constraints on the couplings resulting from the tracer conditions in the disformal frame would result in different conditions than the ones we work with.

To do so, we consider the full EFT action (without having used the Ricci-flatness condition yet) and perform field redefinitions order-by-order.\\
Starting from the order-four action in Eq.~\eqref{eq: quartic correction}, the disformal transformation needed to remove the Ricci-terms at quadratic order reads 
\begin{equation}
    \Tilde{g}_{\mu \nu}^{(4)}=g_{\mu \nu}+\delta g_{\mu \nu}^{(4)}=g_{\mu \nu}+\frac{2a_1+a_2}{2\Lambda_{\rm UV}^2}R\, g_{\mu \nu}-\frac{a_2}{\Lambda_{\rm UV}^2}R_{\mu \nu}\,.
\end{equation}
The action up to level four in derivatives is then altered in the following way
\begin{equation}
\begin{aligned}
  S(\Tilde{g}^{(4)})&=S_{EH}+S_4+\frac{\delta S_{EH}}{\delta g_{\mu \nu}^{(4)}} \delta g_{\mu \nu}^{(4)}+\frac{\delta S_{4}}{\delta g_{\mu \nu}^{(4)}} \delta g_{\mu \nu}^{(4)}\, +\, \text{higher orders}\\
  &=S_{EH}+S_4-\frac{M_{\rm pl}^2}{2}\int \mathrm{d}^4x\sqrt{-g}\left(\frac{a_1}{\Lambda_{\rm UV}^2}R^2+\frac{a_2}{\Lambda_{\rm UV}^2}R_{\mu \nu}R^{\mu \nu} \right)+\frac{\delta S_{4}}{\delta g_{\mu \nu}^{(4)}} \delta g_{\mu \nu}^{(4)} \\
  & \quad \, + \,\text{higher orders}\,,
  \end{aligned}
\end{equation}
where we do not write out the higher order terms as they all include even powers of Ricci-tensors coming from $\delta g_{\mu \nu}^{(4)}$ and thus do not contribute on Ricci-flat background.
Generally, the effect of the field redefinition carries over to order six in derivatives. However, the equations of motion at order four scale as 
\begin{equation}
    \frac{\delta S_{4}}{\delta g_{\mu \nu}^{(4)}} \propto \text{Ric}\,,
\end{equation}
resulting in 
\begin{equation}
    \frac{\delta S_{4}}{\delta g_{\mu \nu}^{(4)}} \delta g_{\mu \nu}^{(4)}\propto \text{Ric}^2\,.
\end{equation}
This means that the order six in derivatives terms with coefficients $b_1$ and $b_2$ are not touched by the field redefinition. Accounting for field redefinitions at order six in derivatives, note that 
\begin{equation}
    \frac{\delta S_{6}}{\delta g_{\mu \nu}^{(6)}} \delta g_{\mu \nu}^{(6)}=O\left(\frac{1}{\Lambda_{\rm UV}^8}\right)\,,
\end{equation}
while higher order corrections come with even higher powers of $1/\Lambda_{\rm UV}$.
This means that our order eight in derivatives couplings are not altered by field redefinitions, either. To sum it up, the values of the couplings associated with terms without Ricci-tensors, $b_1, c_1,c_2,c_3$ are independent of field redefinitions.

\subsection{Comparison to string-theoretic Effective Field Theory coefficients}
The pure gravity string-theoretic EFT up to order eight in derivatives in four dimensions reads~\cite{Melville:2024zjq, Bern:2021ppb, Kawai:1985xq}
\begin{equation}
\begin{aligned}
    S_{\text{ST}}=\frac{M_{Pl}^2}{2}\int \mathrm{d}^4x \sqrt{-g}\Bigg( & R+\frac{\Tilde{b}_1}{\Lambda_{\rm UV}^2}R^{\alpha \beta}_{\,\,\,\,\,\gamma \delta}R^{\gamma \delta}_{\,\,\,\,\,\eta \rho}R^{\eta \rho}_{\,\,\,\,\,\alpha \beta}+\frac{\Tilde{c}_1}{\Lambda_{\rm UV}^4}\left( R_{\alpha \beta\gamma \delta} R^{\alpha \beta\gamma \delta}\right)^2 \\ 
    &+\frac{\Tilde{c}_2}{\Lambda_{\rm UV}^4}\left(\frac{1}{2} R^{\alpha \beta\gamma \delta}\varepsilon_{\alpha \beta}^{\,\,\,\,\,\,\,\rho \sigma} R_{\rho \sigma\gamma \delta}\right)^2 \Bigg)\,.
    \end{aligned}\label{eq: pure gravity superstring}
\end{equation} 
Note that \eqref{eq: pure gravity superstring} uses a basis of curvature scalars containing a dual Riemann tensor term. Even though this term vanishes in spherical symmetry, it needs to be considered as it translates into a linear combination of our basis at order eight, namely
\begin{equation}
\begin{aligned}
    \left(\frac{1}{2} R^{\alpha \beta\gamma \delta}\varepsilon_{\alpha \beta}^{\,\,\,\,\,\,\,\rho \sigma} R_{\rho \sigma\gamma \delta}\right)^2&=\left(R_{\alpha\beta\gamma\delta}R^{\alpha\beta\gamma\delta}\right)^2-4\, {R}_{\quad \mu \nu}^{\kappa \lambda} R^{\mu \nu}_{\quad \rho \sigma}R^{\rho \sigma}_{\quad \tau \omega}R^{\tau \omega}_{\quad \kappa \lambda}\\&\quad+\text{Ricci-terms}\,,
\end{aligned}
\end{equation}
for Lorentzian signature. For Euclidean signature, there is a sign flip in the identity. With this, \eqref{eq: pure gravity superstring} becomes
\begin{equation}
\begin{aligned}
    S_{\text{ST}}=\frac{M_{Pl}^2}{2}\int \mathrm{d}^4x \sqrt{-g} \Bigg( & R+\frac{\Tilde{b}_1}{\Lambda_{\rm UV}^2}R^{\alpha \beta}_{\,\,\,\,\,\gamma \delta}R^{\gamma \delta}_{\,\,\,\,\,\eta \rho}R^{\eta \rho}_{\,\,\,\,\,\alpha \beta}+\frac{\Tilde{c}_1+\Tilde{c}_2}{\Lambda_{\rm UV}^4}\left( R_{\alpha \beta\gamma \delta} R^{\alpha \beta\gamma \delta}\right)^2 \\ 
    &-\frac{4 \, \Tilde{c}_2}{\Lambda_{\rm UV}^4}{R}_{\quad \mu \nu}^{\kappa \lambda} R^{\mu \nu}_{\quad \rho \sigma}R^{\rho \sigma}_{\quad \tau \omega}R^{\tau \omega}_{\quad \kappa \lambda} \Bigg)\,,
    \end{aligned}
\end{equation} 
and we can identify
\begin{equation}\label{eq: translation ST basis to our basis}
    \Tilde{b}_1=b_1\,,\quad \Tilde{c}_1+\Tilde{c}_2=c_1\,, \quad -4\, \Tilde{c}_2=c_2\,.
\end{equation}
For the superstring, the EFT coefficients were shown to be \cite{Melville:2024zjq, Bern:2021ppb, Kawai:1985xq}
\begin{equation}
    \Tilde{b}_1=0\,, \quad \Tilde{c}_1=\Tilde{c}_2\approx0.3\,.
\end{equation}
This lies within our polygon of constraints at order eight but is not compatible with the surface-gravity constraint at order six.

In the case of a massless scalar field coupled to gravity, the most general EFT with massless scalar fields coupled to gravity to order six in derivatives is \cite{Cano:2019ore}
\begin{equation}
\begin{aligned}
S_{\text{ST}}^\phi= & \frac{M_{Pl}^2}{2} \int d^4 x \sqrt{-g}\left(R+\frac{\alpha_1}{\Lambda_{S\rm UV}^2} \phi_1 R_{\alpha \beta \gamma \delta}R^{\alpha \beta \gamma \delta} \right.\\& \left.\hspace{3cm}+\frac{\alpha_2}{\Lambda_{\rm UV}^2}\left(\phi_2 \cos \theta_m+\phi_1 \sin \theta_m\right)  R_{\mu \nu \rho \sigma} \tilde{R}^{\mu \nu \rho \sigma}\right. \\
& \left. \hspace{3cm}+ \frac{\lambda_{\mathrm{ev}}}{\Lambda_{\rm UV}^4} R_{\mu \nu}{ }^{\rho \sigma} R_{\rho \sigma}{ }^{\delta \gamma} R_{\delta \gamma}{ }^{\mu \nu}+ \frac{\lambda_{\mathrm{odd}}}{\Lambda_{\rm UV}^4} R_{\mu \nu}{ }^{\rho \sigma} R_{\rho \sigma}{ }^{\delta \gamma} \tilde{R}_{\delta \gamma}{ }^{\mu \nu} \right.\\& \left.\hspace{3cm}-\frac{1}{2}\left(\partial \phi_1\right)^2-\frac{1}{2}\left(\partial \phi_2\right)^2\right)\,.
\end{aligned}\label{eq: ST EFT with scalar field}
\end{equation}
In the case of the heterotic superstring, the couplings read \cite{Cano:2019ore,Cano:2021rey}
\begin{equation} \label{eq: heterotic superstring scalar}
\lambda_{\mathrm{ev}}=\lambda_{\mathrm{odd}}=\theta_m=0, \alpha_2=-\alpha_1=1 / 8\,.
\end{equation}
For the Einstein-Dilaton-Gauss-Bonnet case, the couplings are \cite{Cano:2019ore}
\begin{equation} \label{eq: einstein-dilaton-gauß-bonnet}
\lambda_{\mathrm{ev}}=\lambda_{\mathrm{odd}}=\alpha_2=0\,.
\end{equation}
Note, that applying \eqref{eq: ST EFT with scalar field} to our spherically symmetric static black hole example, the action reduces to
\begin{equation}
\begin{aligned}
S_{\text{ST}}^\phi= & \frac{M_{Pl}^2}{2}\int d^4 x \sqrt{-g}\left(R+\frac{\alpha_1}{\Lambda_{\rm UV}^2}  \phi_1 R_{\alpha \beta \gamma \delta}R^{\alpha \beta \gamma \delta} \right. \\& \left. \hspace{3cm}+ \frac{\lambda_{\mathrm{ev}}}{\Lambda_{\rm UV}^4} R_{\mu \nu}{ }^{\rho \sigma} R_{\rho \sigma}{ }^{\delta \gamma} R_{\delta \gamma}{ }^{\mu \nu}-\frac{1}{2}\left(\partial \phi_1\right)^2\right)\,,
\end{aligned}\label{eq: ST EFT with scalar field 2}
\end{equation}
such that we can only make statements about $\alpha_1=b_3\,, \, \lambda_{\mathrm{ev}}=b_1$. The heterotic superstring prediction \eqref{eq: heterotic superstring scalar} as well as the Einstein-Dilaton-Gauss-Bonnet prediction \eqref{eq: einstein-dilaton-gauß-bonnet} allow to meet all constraints but the proper time constraint simultaneously, compare Fig.~\ref{fig: couplings 6 with scalar hair}.  \\
We have found that our full set of regularity constraints are compatible with string theoretical predictions when imposing regularity starting from order eight in derivatives.

\subsection{Comparison to Effective Field Theory coefficient from asymptotic safety}
An analysis of a correction to the Einstein-Hilbert action by the Goroff-Sagnotti counter-term
\begin{equation} \label{eq: ASQG action}
\begin{aligned}
    S_{AS}^{\text{eucl.}}&=\int \mathrm{d}^4x \sqrt{g} \left( - \frac{1}{16 \pi G_N}(R-2 \Lambda)+ B_1^{\text{eucl.}}\, C_{\mu\nu }^{\ \ \rho\sigma }C_{\rho\sigma }^{\ \ \delta\gamma }C_{\delta\gamma }^{\ \ \mu\nu } \right)
    \end{aligned}
\end{equation}
within the framework of asymptotically safe quantum gravity leads to an interacting
fixed point in Euclidean signature located at \cite{Baldazzi:2023pep}
\begin{equation} \label{eq: GFP couplings}
g_{N,\,\ast}=k^2 G_{N\,\ast}=0.707,\quad \lambda_{\ast} = \Lambda_{\ast}/k^2 = 0.193,\quad  b_{1\,\ast}^{\rm eucl.}=k^2B_{1,\ast}^{\rm eucl.}= 1.37 \cdot 10^{-4},
\end{equation}
where $k$ is the FRG scale, acting as a cutoff scale in the infrared. The dimensionful couplings are denoted by uppercase and the dimensionless couplings by lowercase letters. The overall sign in \eqref{eq: ASQG action} is chosen in order to have a negative sign in front of $S_{AS}^{\text{eucl.}}$ in the path integral. As explained in \ref{app: weyl basis}, the coupling in front of the $C_{\mu\nu }^{\ \ \rho\sigma }C_{\rho\sigma }^{\ \ \delta\gamma }C_{\delta\gamma }^{\ \ \mu\nu }$ term corresponds to the coupling in front of the $R_{\mu\nu }^{\ \ \rho\sigma }R_{\rho\sigma }^{\ \ \delta\gamma }R_{\delta\gamma }^{\ \ \mu\nu }$ term, as we switch from the Weyl curvature basis to the Riemann curvature basis. The expressions in this section all refer to Euclidean signature calculations. Starting from the fixed point, 
the value of the coupling in the EFT can be obtained from an integration to $k \rightarrow 0$, see, e.g., \cite{Knorr:2024yiu,Eichhorn:2024wba,Eichhorn:2026inw} for examples, with $G_N = \rm const$ for $k> M_{\rm Planck}$ and gravitational fluctuations vanishing below. Thus $B_1^{\rm eucl.}>0$ is expected to come out.

Naively, one could expect a Wick rotation to introduce an additional minus sign to the coupling in Lorentzian signature, coming from the transformation of the volume element. By a naive Wick rotation, the fully contracted Weyl-cubed term does not introduce any signs. This can be understood by restricting ourselves to metrics of the type 
\begin{equation}
     \dd s^2 =\pm h(r) \dd t^2 + \frac{1}{f(r)}\, \dd r^2 + r^2 \mathrm{d} \Omega^2\,,
\end{equation}
which importantly is time-independent and diagonal and comparing the Weyl-cubed term to  its Euclidean counterpart. However, a full Lorentzian analysis of the problem is required for a final assessment. 
\subsection{Comparison to positivity bounds}
Causality considerations constrain the couplings in \eqref{eq: pure gravity superstring} \cite{deRham:2021bll}. 
There are two cases to distinguish, depending on the value of the cutoff of the EFT. For the first case, for which the cutoff is above $10^{-9}\,\rm eV$,  no constraint on $b_1$ exists. For the second case, for which the cutoff is below that value, a constraint on $b_1$ exists and requires $b_1 \geq 0$.
Our $\kappa$ constraint at order six gives $b_1>0$, which is compatible with the causality constraint, whenever it can be imposed. 

In contrast, the constraints on order-eight couplings do not require any conditions on the cutoff scale. These conditions are 
\begin{equation}
\Tilde{c}_1>0\,, \quad \Tilde{c}_2>0\,,
\end{equation}
which, using \eqref{eq: translation ST basis to our basis}, corresponds to
\begin{equation}
 4\,c_1+c_2>0\,, \quad c_2<0\,.
\end{equation}
The first inequality is identical to our $\kappa$ constraint at order eight in derivatives and $c_2<0$  also lies within the polygon of combined constraints.

Overall we find that also the causality considerations are compatible with our regularity constraints, but that regularity imposes constraints that hold more generally.

We consider it noteworthy that positivity bounds -- when they constrain the couplings -- are compatible with our constraint of \emph{increased}, rather than decreased surface gravity. A priori, one may have considered decreased surface gravity as another sign of weakened gravity. This would, as we have argued, weaken the instability of the Cauchy horizon, which must necessarily be there to avoid the existence of a stable spacetime in which predictability is lost at the Cauchy horizon. Therefore, we have imposed increased surface gravity and find it noteworthy that a different line of reasoning, also tied to the consistency of time-evolution in the theory leads to the same constraint.

\section{Conclusions and outlook} \label{sec: conclusion}
The landscape of gravitational theories beyond GR is vast. It may become even vaster, if hints that a restriction to second-order equations of motion is unwarranted to guarantee physically acceptable behavior of solutions \cite{Held:2025fii}, can be strengthened. Invoking quantum gravity is, as of yet, not sufficient to reduce this landscape by much. This is partially because one well-developed candidate framework, string theory, itself comes with a vast landscape of low-energy effective field theories that it can UV complete. A second framework, asymptotically safe quantum gravity, is expected to have a significantly smaller landscape \cite{Eichhorn:2026uqj}. However, the calculation of the effective action in asymptotically safe quantum gravity is not yet advanced enough to know the couplings with sufficient precision and in Lorentzian signature, although important steps towards understanding the spacetime structure of black holes in asymptotic safety  are being made \cite{DelPorro:2025wts}. Against this background, additional principles to reduce the landscape of viable gravitational theories are called for. Here, we focus on physically acceptable black-hole spacetimes, together with the EFT formulation of gravitational theories, as a principle to achieve such a reduction of the landscape.

Black holes in GR are physically unacceptable spacetimes, even if their exterior spacetime geometries provide adequate predictions for experiments at currently achievable precision. Their inconsistencies, which consist in curvature singularities, geodesic incompleteness and breakdown of the initial-value problem at Cauchy horizons, are not automatically resolved within a modified gravitational action. Thus, we expect that, within the general space of modified gravitational theories, spanned, e.g., by all possible couplings of all possible gravitational degrees of freedom, there is a bounded region that contains black-hole spacetimes without such artefacts. However, the absence of these artefacts and consistency with observational data does not, in our view, automatically make them plausible candidates for physically acceptable black-hole spacetimes. For instance, we consider it implausible that beyond-GR-effects are relatively larger far outside the horizon than they are in the black-hole interior. 
Such considerations should further limit the allowed modified gravitational theories. 

Specifically, we make two assumptions: First, we assume that the resolution of curvature singularities and geodesic incompleteness relies on a suitable ``weakening of gravity'' and that this weakening is present throughout the spacetime (although with decreasing magnitude at larger distances from the black-hole center). This simplicity or monotonicity assumption enables us to connect regularity of the black-hole interior with properties of the black-hole exterior. \\
Second, we assume that these physical requirements are already achieved in an EFT expansion of the full theory. This encodes the expectation that the above physical effects ``build up'' order by order in the EFT expansion.\\
Both of these assumptions -- the existence of tracers of regularity throughout a black-hole spacetime and their visibility already in the EFT expansion -- can be made more concrete in distinct ways; e.g., distinct tracers can be formulated and the EFT can contain additional light fields besides gravity. On this basis, distinct ``islands'' in the EFT can be identified and we consider the ``island'' that we delineate a particularly plausible, but very likely not the only ``island''.\\

Therefore, our goals are to i) formulate precise conditions for ``weakening of gravity'' and resolution of Cauchy horizons and ii) establish at which order of the EFT expansion all conditions can for the first time be met. 

The resulting, specific ``island of less-singular black holes'' singles out a finite, bounded region in the EFT's Wilson coefficients. We find that for an EFT with only the metric as a light gravitational degree of freedom, this island comes into existence at order eight in the EFT. We highlight that it is non-trivial that this island exists, because there are only four independent Wilson coefficients at order eight, when restricted to spherically symmetric, static spacetimes, to realize our five conjectures. The existence of a bounded region -- a regular-black-hole island -- is therefore a significant finding.

 For an EFT with an additional light scalar, at order six, despite the added coupling pertaining to the scalar, is also insufficient for the island to exist. Islands realizing various subsets of the five conjectures exist.

Wilson coefficients outside this island are part of the ``swamp'' in that black-hole solutions are not less-singular. This does not strictly invalidate such Wilson coefficients, but makes them -- in our view -- less plausible. We expect that the physical mechanisms at the heart of regularizing black holes leave perturbative imprints, such that they are visible already in the EFT expansion. We also expect our conditions on the spacetimes themselves to be plausible, because they exclude, e.g., large changes to gravitational effects that would correspond to ``stronger gravity'' far outside a black-hole horizon. However, under the monotonicity assumption made above, it is possible to formulate distinct sets of conditions that constrain the EFT coefficients. For example, we may slightly modify some of our requirements, e.g., when we consider the proper time for infalling observers, non-radial motion may be considered, or different physical initial and final points of the motion in the spacetime may be considered. This modifies the particular island that we investigate and there may well be an ``archipelago'' of closely related islands.

More broadly, different such islands can be found, which correspond to different physical conditions on the effective field theory. Delineating these islands and understanding their relation to each other enables us to learn about the physical content of the Wilson coefficients, irrespective of whether any given set of conditions is assumed to be most plausible. As such, our construction principle, based on the monotonicity (simplicity) assumption and the EFT expansion, is more generally applicable.

Importantly, our specific results constrain the Wilson coefficients independently of known positivity bounds, thereby providing genuinely new insights. For instance, positivity bounds have so far been unable to constrain the sign of the dimension-6 operator $\sim \mathrm{Riem}^3$, without making strong assumptions about the cutoff of the EFT. In contrast, our conjectures single out $b_1>0$.

We also make a comparison with calculations of Wilson coefficients in candidate quantum theories of gravity. For asymptotically safe quantum gravity, this comparison is largely aspirational, because a constraint on the Wilson coefficient exists, but only in the Euclidean setting, calling for a corresponding Lorentzian calculation for a truly meaningful comparison. For string theory, the superstring calculations in \cite{Melville:2024zjq, Bern:2021ppb, Kawai:1985xq} give rise to Wilson coefficients within our ``less-singular island'' at order eight in the EFT. This result may be interpreted as an encouragement for the viability of our considerations in that at least one candidate theory of quantum gravity appears to be compatible with them.\\

As an outlook, let us highlight that singularity resolution in static, spherically symmetric black-hole spacetimes may not be physically relevant. After all, all existing black holes in our universe are expected to have formed from gravitational collapse and are not eternal. Therefore, it may be physically more relevant to require similar conditions to the ones that we imposed, but for the case of a time-dependent metric describing gravitational collapse, possibly also loosening the assumption of spherical to axisymmetry. We expect that a similar analysis to the one we have done should in principle be possible. 

In addition, delineating the distinct islands that realize different sets of conditions enables us to understand the physics that is encoded in the Wilson coefficients of the EFT expansion. We consider this an interesting future direction that is complementary to studies of black-hole (like) objects in specific candidate fundamental frameworks.

\subsubsection*{Acknowledgements}
We gratefully acknowledge discussions with and helpful comments on a draft of the paper by Johanna Borissova. We are also grateful to Benjamin Knorr for discussions and insights on the basis of curvature invariants.
A.~E.~and P.~G.~S.~F.~are grateful to the organizers of the workshop ``Crossing over the edge'' at the Institute for the Fundamental Physics of the Universe, Trieste, for the invitation to a very stimulating workshop and thank the participants for very insightful discussions.
A.~E.~acknowledges the European Research Council's (ERC) support under the European Union’s Horizon 2020 research and innovation program Grant agreement No.~101170215 (ProbeQG). This work is partially funded by the Deutsche Forschungsgemeinschaft (DFG, German Research Foundation) under Germany’s Excellence Strategy EXC 2181/1 - 390900948 (the Heidelberg STRUCTURES Excellence Cluster). P.~G.~S.~F.~is supported by national funds through FCT - Fundação para a Ciência e a Tecnologia, I.P., under the Scientific Employment Stimulus - Individual Call - contract 2024.11822.CEECIND. V.~K.~is supported by a scholarship of the Friedrich-Ebert-foundation.

\newpage
\appendix

\section{Constructing the Effective Field Theory}

\subsection{Independent curvature invariants at order eight without field redefinitions}\label{app:constructing_the_eft}
The literature is clear on the most general non-degenerate parity-invariant contributions to the Lagrangian up to order six in derivatives \cite{Fulling:1992vm,Cano:2019ore}. In this section, we discuss how the parity-invariant curvature basis at order eight in derivatives given in \cite{Fulling:1992vm} reduces to \eqref{eq: action order eight Riem basis} under the Ricci-flat condition in four dimensions \textit{without applying field redefinitions}.  
For ease of comparison with the original work \cite{Fulling:1992vm}, we deviate from the convention used in the main text of our paper and work with the Latin index labeling used in the original work. \\
We consider the \textit{non-derivative} curvature invariants first. By that, we mean curvature invariants only consisting of contractions of curvature tensors, and not containing any additional covariant derivatives. Generally, in d dimensions, there are 26 such terms \cite{Fulling:1992vm}. Considering Ricci-flatness, only three of them remain in $d=4$. Namely,
\begin{align}
    &T_A=R R^{pqrs}R_{p \, \, r}^{\, \,\, t \, \, \, u}R_{qtsu}\,,\\
    &T_B=\left(R^{abcd}R_{abcd}\right)^2\,,\\
    &T_C= R^{pqrs}R_{pq}^{ \, \, \, \, \, tu}R_{tu}^{ \, \, \, \,vw}R_{rswv}\,.
\end{align}
Using dimensional constraints, $T_A$ can be rewritten as
\begin{equation}
\begin{aligned}
   T_A=R R_{q \, \, \, s}^{\, \, \,p\,\,\,r}R_{p \, \, \, r}^{\, \, \,t\,\,\,u}R_{t \, \, \, u}^{\, \, \,q\,\,\,s}&=  \frac{1}{2}R R_{qs}^{\,\,\,\,pr}R_{pr}^{\,\,\,\,tu}R_{tu}^{\,\,\,\,qs}+F(\text{Ric}^2)\,,
    \end{aligned}
\end{equation}
as shown in the appendix of \cite{Cano:2019ore}. Here, $F$ represents a function proportional to Ricci-squared terms. This way, we have found the first three terms in \eqref{eq: action order eight Riem basis}.\\
We proceed with the \textit{derivative} curvature invariants. These are those containing contractions of covariant derivatives of the Riemann curvature tensor. Generally, there are 66 such terms in d dimensions \cite{Fulling:1992vm}. Naively, using only the Ricci-flat condition, ten of these remain in $d=4$. Those are 
\begin{align}\label{eq: derivative terms}
 T_1&=\nabla^t \nabla^s \nabla^r R^{pq}\nabla_tR_{prqs}\,,\\
 T_2&= \nabla^u \nabla^t R^{pqrs}\nabla_u \nabla_t R_{pqrs}\,,\\
T_3&=\Box R R^{abcd}R_{abcd}\,, \\
T_4&=\nabla^s \nabla^r R^{pq} R^{tu}_{\,\,\,\,pr} R_{tuqs}\,,\\
 T_5&= R^{pq}R^{rstu}\nabla_q\nabla_p R_{rstu} \,,\\
 T_6&= \nabla^p R R^{qrst} \nabla_p R_{qrst}\,,\\
 T_7&= R \nabla^t R^{pqrs} \nabla_t R_{pqrs} \,,\\
 T_8&= R^{pq}\nabla_p R^{rstu} \nabla_q R_{rstu} \,,\\
 T_9&= \nabla^r R^{pq} R^{st \, \, \,u}_{\,\,\,\,p} \nabla_u R_{stqr}\,,\\
 T_{10}&= R^{pqrs} \nabla_q R^{tuv}_{\,\,\,\,\,p} \nabla_s R_{tuvr}\,.
\end{align}
In the following, we show that only $T_3$ and $T_9$ need to be considered at the level of the action. Additionally to sucessive partial integration, we can do so by using several identities. The first one is
\begin{equation}
\begin{aligned}
I_1: \quad \nabla^e \nabla_e R_{a b c d}= & 2 \nabla_{[a \mid} \nabla_c R_{\mid b] d}+2 \nabla_{[b \mid} \nabla_d R_{\mid a] c}-4\left[R_{\, \,\, a \, \, \, b}^{p \, \, \, \, q} R_{p[c|q| d]}+R^{p \, \, \, q}_{a \, \, \,[c\mid} R_{p b q \mid d]}\right] \\
& +g^{p q}\left[R_{q b c d} R_{p a}+R_{a q c d} R_{p b}\right]\,,
\end{aligned}
\end{equation}
which was shown to be true in \cite{Andrews:2011}. Secondly, we are using the definition of the Riemann-tensor components in terms of the commutator of two covariant derivatives
\begin{equation}
    I_2: \quad \left[\nabla_a,\nabla_b\right]v^c=R^c_{\,\,dab}v^d\,,
\end{equation}
with an arbitrary vector field $v$.
Finally, we need the contracted as well as the non-contracted differential Bianchi identities
\begin{align}
    &I_3: \quad \nabla^aR_{ab}=\frac{1}{2}\nabla_b R\,, \\
    &I_4: \quad \nabla^aR_{abcd}=\nabla_c R_{bd}-\nabla_d R_{bc}\,,\\
    &I_5: \quad 0=\nabla_eR_{abcd}+\nabla_cR_{abde}+\nabla_dR_{abec}\,.
\end{align}
From now on, all equations are to be thought of as written under an integral, as we are interested in the relevant terms at the level of the EFT action. We will therefore apply partial integration and drop the boundary terms directly. Additionally, we will use the notation $\propto$ for equality up to Ricci-squared terms and use schematic notation whenever helpful. For example, we write "$\text{Riem} \,\text{Ricci}^2$" whenever a term appears for which it is only relevant to know that it is proportional to a Riemann and two Ricci tensors without additional covariant derivatives. Before we start, note that we can rewrite $T_9$ as
\begin{equation}
\begin{aligned}
 T_9=\nabla^r R^{pq} R^{st \, \, \,u}_{\,\,\,\,p} \nabla_u R_{stqr}&=-\nabla_u \nabla^r R^{pq} R^{st \, \, \,u}_{\,\,\,\,p}R_{stqr}-\nabla^r R^{pq} \nabla_uR^{st \, \, \,u}_{\,\,\,\,p}R_{stqr}\\
  &\stackrel{I_4}{\propto}-\nabla_u \nabla^r R^{pq} R^{st \, \, \,u}_{\,\,\,\,p}R_{stqr}\\
  &=- R^{pust}R^{qr}_{\,\,\,\,st}\nabla_u \nabla_rR_{pq}\\
  &=- R^{upst}R^{rq}_{\,\,\,\,st}\nabla_u \nabla_rR_{pq}\,,
  \end{aligned}
\end{equation}
which corresponds to the $c_5$ term in \eqref{eq: action order eight Riem basis}.\\
We start off by considering the terms with four covariant derivatives, namely $T_1$ and $T_2$. We find 
\begin{equation}
    \begin{aligned}
        T_1= \nabla^t \nabla^s \nabla^r R^{pq}\nabla_tR_{prqs}= -\nabla^s \nabla^r R^{pq}\Box R_{prqs}\stackrel{I_1}{\propto}\nabla \nabla \text{Ric}\,\text{Riem}^2\,,
    \end{aligned}
\end{equation}
which reduces to a term containing only two covariant derivatives. This means that $T_1$ is already accounted for by the curvature invariants containing fewer covariant derivatives, and we can drop it without loss of generality. Going on, we find
\begin{equation}
    \begin{aligned}
        T_2&=\nabla^u \nabla^t R^{pqrs}\nabla_u \nabla_t R_{pqrs}=\nabla_t \nabla_u\nabla^u \nabla^t R^{pqrs}R_{pqrs}\\&\stackrel{I_2}{=}\Box \text{Riem}\Box \text{Riem}+\nabla \nabla\left( \text{Riem} \,\text{Riem} \right) \text{Riem}+\nabla \left( \text{Riem} \nabla\text{Riem} \right)\,\text{Riem}\\
        &\stackrel{I_1}{=}\nabla \nabla \text{Ric}\nabla \nabla \text{Ric}+\text{Riem}^4+\text{Riem}^2 \nabla \nabla \text{Ric} +\nabla \nabla\left( \text{Riem}\,\text{Riem} \right) \text{Riem}\\ & \quad+\nabla \left( \text{Riem} \nabla\text{Riem} \right)\,\text{Riem}\\
        &\propto\text{Riem}^4+\text{Riem}^2 \nabla \nabla \text{Ric}+\nabla \nabla\left( \text{Riem} \text{Riem} \right) \text{Riem}+\nabla \text{Riem} \nabla\text{Riem}\,\text{Riem}\,.
    \end{aligned}
\end{equation}
These terms are also already taken into account by the curvature invariants containing less than four covariant derivatives.
We now move on to the curvature invariants containing two covariant derivatives, $T_3$ to $T_{10}$. We start by showing that, under the integral, $T_4$ is proportional to $T_9$:
\begin{equation}
    \begin{aligned}
T_4= \nabla^s \nabla^r R^{pq} R^{tu}_{\,\,\,\,pr} R_{tuqs}&= -\nabla^r R^{pq}\nabla^s R^{tu}_{\,\,\,\,pr} R_{tuqs}-\nabla^r R^{pq} R^{tu}_{\,\,\,\,pr}\nabla^s R_{tuqs}\\
&\stackrel{I_4}{=}-\nabla^r R^{pq}\nabla^s R^{tu}_{\,\,\,\,pr} R_{tuqs}-\nabla^r R^{pq} R^{tu}_{\,\,\,\,pr}\left(\nabla_u R_{qt} -\nabla_t R_{qu}\right)\\
&\propto -\nabla^r R^{pq}\nabla^s R^{tu}_{\,\,\,\,pr} R_{tuqs}\\
&=-\nabla^r R^{pq} R^{st \, \, \,u}_{\,\,\,\,p} \nabla_u R_{stqr}=-T_9\,,
    \end{aligned}
\end{equation}
where we have relabeled the indices and used the symmetry of the Ricci tensor in the last step.
For $T_6$, we find
\begin{equation}
\begin{aligned}
        T_6= \nabla^p R R^{qrst} \nabla_p R_{qrst}&=-R \nabla^p  R^{qrst} \nabla_p R_{qrst}- R  R^{qrst} \Box R_{qrst}\\
        &\stackrel{I_1}{\propto}-R \nabla^p  R^{qrst} \nabla_p R_{qrst}-R\, \text{Riem}^3\\
         &=-T_7-R\, \text{Riem}^3\,,
\end{aligned}
\end{equation}
meaning that it turns out to consist of a non-derivative contribution, that is already accounted for by $T_A,\,T_B,\,T_C$, plus a term proportional to $T_7$. Analyzing $T_7$, we find that, up to terms that are already accounted for by $T_A, \, T_B, \, T_C$, it is identical to $T_3$ up to rescaling:
\begin{equation}
    \begin{aligned}
        T_7=  R \nabla^t R^{pqrs} \nabla_t R_{pqrs}&=- \nabla_tR \nabla^t R^{pqrs}  R_{pqrs}- R \Box R^{pqrs}  R_{pqrs}\\
         &\stackrel{I_1}{\propto}- \nabla_tR \nabla^t R^{pqrs}  R_{pqrs}+\text{Ric}\, \text{Riem}^3\\
         &= \frac{1}{2}\Box R  R^{pqrs}  R_{pqrs}+\text{Ric}\, \text{Riem}^3\\
          &= \frac{1}{2}T_3+\text{Ric}\, \text{Riem}^3\,.
    \end{aligned}
\end{equation}
For $T_8$, we get
\begin{equation}
    \begin{aligned}
T_8&= R^{pq}\nabla_p R^{rstu} \nabla_q R_{rstu}\\
&=-\nabla_qR^{pq}\nabla_p R^{rstu}  R_{rstu}-R^{pq}\nabla_q\nabla_p R^{rstu}  R_{rstu}\\&=-\frac{1}{2}\nabla_qR^{pq}\nabla_p \left(R^{rstu}  R_{rstu} \right)-T_5
\\&
\stackrel{I_3}{=}\frac{1}{4}\Box R \left(R^{rstu}  R_{rstu} \right)-T_5=\frac{1}{4}T_3-T_5\,.
    \end{aligned}
\end{equation}
The term $T_{10}$ is identical to $\mathcal{L}_1$ in \cite{Bueno:2019ltp}, where it is shown that
\begin{equation}
    \begin{aligned}
        T_{10}&= R^{pqrs} \nabla_q R^{tuv}_{\,\,\,\,\,\,\,\,p} \nabla_s R_{tuvr}=-\frac{1}{8}R^{abcd}\Box R^{ef}_{\,\,\,\,\,ab}R_{efcd}\\
         &\stackrel{I_3}{=}\text{Riem}^4-\frac{1}{8}R^{abcd} R^{ef}_{\,\,\,\,cd} \left(\nabla_e \nabla_a R_{fb}-\nabla_f \nabla_a R_{eb}+\nabla_f \nabla_b R_{ea}-\nabla_e \nabla_b R_{fa}\right)  \\
         &=\text{Riem}^4-\frac{1}{2}R^{abcd} R^{ef}_{\,\,\,\,cd} \nabla_e \nabla_a R_{fb} \\
         &=\text{Riem}^4-\frac{1}{2}T_9\,,
    \end{aligned}
\end{equation}
where we have used the antisymmetry of the first two Riemann tensors under $a\leftrightarrow b$ and $e \leftrightarrow f$, respectively. Finally, we still need to consider $T_5$:
\begin{equation}
\begin{aligned}
     T_5&= R^{pq}R^{rstu}\nabla_q\nabla_p R_{rstu}\\
     &\stackrel{I_5}{=}- R^{pq}R^{rstu}\nabla_q\left( -\nabla_r R_{tups}+\nabla_s R_{tupr}\right)\\
     &=2R^{pq}R^{rstu}\nabla_q\nabla_r R_{tups}\\
     &\stackrel{I_2}{=}2R^{pq}R^{rstu}\nabla_r\nabla_q R_{tups}+ \text{Ric}\,\left(\text{Riem}\right)^3\\
     &\stackrel{I_4}{\propto}-2\nabla_rR^{pq}R^{rstu}\nabla_q R_{tups}+\text{Ric}\,\left(\text{Riem}\right)^3\\
     &=\underbrace{2\nabla_q\nabla_rR^{pq}R^{rstu} R_{tups}}_{:=S_1}+\underbrace{2\nabla_rR^{pq}\nabla_qR^{rstu} R_{tups}}_{:=S_2}+\text{Ric}\,\left(\text{Riem}\right)^3\,.    
\end{aligned}
\end{equation}
Consider the first part, $S_1$:
\begin{equation}
\begin{aligned}
S_1&=2\nabla_q\nabla_rR^{pq}R^{rstu} R_{pstu}\\
 &\stackrel{\eqref{eq: Weyl identity 4dim}}{\propto}\frac{1}{2}\nabla_q\nabla^rR^{pq}g_{rp} R^{abcd}R_{abcd}\\
&=\frac{1}{2}\nabla_q\nabla_pR^{pq} R^{abcd}R_{abcd}\\
&\stackrel{I_3}{=}\frac{1}{4}\Box R R^{abcd}R_{abcd}=\frac{1}{4}T_3\,.
\end{aligned}
\end{equation}
For the second part, $S_2$, we find:
\begin{equation}
    \begin{aligned}
        S_2&=2\nabla_rR^{pq}\nabla_qR^{turs} R_{tups}\\
        &\stackrel{I_5}{=}2\nabla_rR^{pq}\nabla^r R^{tu\,\,\,s}_{\,\,\,\,\,q} R_{tups}-2\nabla_rR^{pq}\nabla^s R^{tu\,\,\,r}_{\,\,\,\,\,q}  R_{tups}\\
        &=\nabla_rR^{pq}\nabla^r \left(R^{\,\,\,stu}_{q} R_{pstu}\right)-2\nabla_rR^{pq}\nabla^s R^{\,\,\,rtu}_{q}  R_{tups}\\
         &=\nabla_rR^{pq}\nabla^r \left(R^{\,\,\,stu}_{q} R_{pstu}\right)-2\nabla^rR^{pq}\nabla^u R_{stqr}  R^{st}_{\,\,\,pu}\\
         &=\nabla_rR^{pq}\nabla^r \left(R^{\,\,\,stu}_{q} R_{pstu}\right)-2T_9\\
         & \stackrel{\eqref{eq: Weyl identity 4dim}}{\propto}-2T_9+ \frac{1}{4}\nabla_r R\nabla^r \left(R_{abcd}R^{abcd}\right)\\
         &=-2T_9- \frac{1}{4}T_3\,.
    \end{aligned}
\end{equation}
Altogether, this means
\begin{eqnarray}
    I_5=\frac{1}{4}T_3-2T_9-\frac{1}{4}T_3+ \text{Ric} \, \text{Riem}^3=-2T_9+ \text{Ric} \, \text{Riem}^3\,.
\end{eqnarray}
With this, we have shown that at order eight in derivatives, the only terms that need to be considered in the EFT action in $d=4$ under the conditions of Ricci-flatness and parity-invariance are those used in \eqref{eq: action order eight Riem basis}.

\subsection{Formulation in the Weyl curvature basis} \label{app: weyl basis}
In order to easily compare our results to  other settings, e.g., calculations in asymptotic safety \cite{Baldazzi:2023pep}, we also provide the rewriting of the EFT action in terms of the Weyl tensor $C$ and the traceless Ricci tensor $S$. At order six in derivatives, this means
\begin{equation}
\begin{aligned}
    \mathcal{L}_6&=b_1 C_{\mu\nu }^{\ \ \rho\sigma }C_{\rho\sigma }^{\ \ \delta\gamma }C_{\delta\gamma }^{\ \ \mu\nu }+\left(\frac{b_1}{2}+b_2\right) C_{\mu\nu\rho\sigma }C^{\mu\nu\rho\sigma }R+6b_1 S^{\alpha \beta} C_{\alpha}^{\, \, \mu \nu \rho}C_{\beta\mu \nu \rho}\\ 
    &=\left(\frac{b_1}{2}+b_2\right) C_{\mu\nu\rho\sigma }C^{\mu\nu\rho\sigma }R+b_1 C_{\mu\nu }^{\ \ \rho\sigma }C_{\rho\sigma }^{\ \ \delta\gamma }C_{\delta\gamma }^{\ \ \mu\nu }\,,
\end{aligned}
\end{equation}
where we used that, due to the Weyl identity in four dimensions
\begin{equation}\label{eq: Weyl identity 4dim}C_{\alpha}^{\, \, \gamma \delta \eta}C_{\beta \gamma \delta \eta}=\frac{1}{4}g_{\alpha \beta} C^{\gamma \delta \eta \mu}C_{\gamma \delta \eta \mu}\,, \end{equation}
the second to last term in the first line vanishes
\begin{equation}
    \left( S^{\alpha \beta} C_{\alpha}^{\, \, \mu \nu \rho}C_{\beta\mu \nu \rho}\right)|_{d=4}=0\,.
\end{equation}
At order eight in derivatives, the Lagrangian becomes
\begin{equation}
    \begin{aligned}\label{eq: action order eight}
        \mathcal{L}_8
        &=c^{(C)}_1 (C_{\alpha \beta \gamma \delta} C^{\alpha \beta \gamma \delta})^2+c^{(C)}_2 {C}_{\quad \mu \nu}^{\kappa \lambda} C^{\mu \nu}_{\quad \rho \sigma}C^{\rho \sigma}_{\quad \tau \omega}C^{\tau \omega}_{\quad \kappa \lambda}+c^{(C)}_3\, R \, C^{\mu \nu}_{\quad \rho \sigma}C^{\rho \sigma}_{\quad \tau \omega}C^{\tau \omega}_{\quad \mu \nu}\\&\,\,\,\,\,\,\,+c^{(C)}_4 C_{\mu \nu \rho \sigma} C^{\mu \nu \rho \sigma}\, \Box\, R+c^{(C)}_5 C^{\mu \nu \alpha \beta}C^{\rho \sigma}_{\quad \alpha \beta}\nabla_{\mu}\nabla_{\rho} S_{\nu \sigma}\,.
    \end{aligned}
\end{equation}
The coefficients of the Riemann and the Weyl curvature basis are related by
\begin{align}
    c^{(C)}_1=c_1\,, \,c^{(C)}_2=c_2\,, \,c^{(C)}_3=\frac{2c_2}{3}+c_3\,, \,c^{(C)}_4=c_4+\frac{c_5}{16}\,,\, c^{(C)}_5=c_5\,,
\end{align}
 for which we have used the Weyl identity \eqref{eq: Weyl identity 4dim}, as well as
 \begin{equation}
     \left( S^{\alpha \beta} C_\alpha^{\,\, \gamma \delta \eta} C_{\beta \gamma}^{\,\,\,\, \,\rho \mu} C_{\delta \eta \rho \mu}\right)|_{d=4}=0\,.
 \end{equation}
\bibliographystyle{JHEP}

\bibliography{Bibliography.bib}

\end{document}